\documentclass[a4paper]{report}
\usepackage[babel, en]{ku-forside}
\usepackage[top=3.5cm, bottom=4cm, left=3.1cm, right=3.1cm, headsep=14pt]{geometry}
\usepackage{amsmath}
\usepackage{subfig}
\usepackage{wrapfig}

\usepackage[english]{babel}
\usepackage[utf8x]{inputenc}
\usepackage{amsmath}
\usepackage{graphicx}
\usepackage{subfig}
\usepackage{pdfpages}
\usepackage{float}
\usepackage{hyperref}
\usepackage{fancyhdr}
\usepackage[para,online,flushleft]{threeparttable}
\usepackage{graphicx}
\usepackage{listings}             
\usepackage{hyperref}

\definecolor{light-gray}{gray}{0.95}

\DeclareMathAlphabet      {\mathbfit}{OML}{cmm}{b}{it}

\let\oldsqrt\sqrt
\def\sqrt{\mathpalette\DHLhksqrt}
\def\DHLhksqrt#1#2{%
    \setbox0=\hbox{$#1\oldsqrt{#2\,}$}\dimen0=\ht0
    \advance\dimen0-0.2\ht0
    \setbox2=\hbox{\vrule height\ht0 depth -\dimen0}%
{\box0\lower0.4pt\box2}}

\titel{Protein Structure Determination } %
\undertitel{} %
\opgave{PhD Thesis} 
\forfatter{Anders S. Christensen}%
\dato{\today}%
\vejleder{Academic supervisor: Jan H. Jensen} 

\begin{document}
\pagenumbering{roman}
\maketitle
\chapter*{Acknowledgements}
\addcontentsline{toc}{chapter}{Acknowledgements}
This thesis represents the work I have carried out as a PhD student under the supervision of Professor Jan H. Jensen in his group of Biocomputational Chemistry.
Thank you to all who have supported me during my work at the third floor of C-building at the H.C. Ørsteds Institute.
\\\\I would especially like to thank the following people:

\begin{itemize}
\item Thank you to my supervisor, Jan "Yoda" Jensen for introducing me to the
    exciting fields of quantum chemistry and biocomputational chemistry, teaching me everything I know (and more), for your patience, and the inspiration you bring to everyone around you.

\item Thank you to Jens Breinholt at Novo Nordisk for supporting me with my work -- I sincerely hope, that my work will very soon become practically useable. And thank you to the Novo Nordisk STAR PhD program for financial support, and for giving me the opportunity to carry out this study.

\item Thank you to all of our collaborators at the Biocenter, who always have been very supportive. Especially, Thomas Hamelryck (in the presence of whom everything is trivially solved using Bayes' theorm) for helping me out with Bayesian theory, your great ideas and more, Wouter Boomsma for being seemingly all-knowing in what concerns PHAISTOS and always being exceptionally helpful, Simon Olsson for helping out with the implementation of the Jeffrey's prior code, and Kresten Lindorff-Larsen for always being encouraging and sharing your knowledge in this field.

\item Thank you to my office mates, Casper Steinmann, Jimmy Kromann and Lars Bratholm, for the invaluable company, our office pranks, and endless number of energy drinks consumed, as well as the highly valuable scientific discussions we continue to share daily (not forgetting the virtual monster we've slayed).

Also thank you to those close colleagues who came by for coffee and friendly conversations; Jonas Elm, Jacob Lykkebo, Nini Reeler, Frederik Beyer (and many, many more!).

\item Thank you to everyone at the Department of Chemistry, especially Kurt V.~Mikkelsen, Stephan P.~A.~Sauer and Sten Rettrup for always being so helpful with everything from bureaucratic procedures, to coupled cluster theory, to derivation of the Slater-Kloster tables.

\item Thank you to all the students in the courses I've taught, and especially the very talented students who have carried out Master's, Bachelor's and various research projects under my supervision. Of those not already mentioned (and in no particular order): Maher Channir, Anders Larsen, Rie Nielsen, Christine Skibsted, Cecilie Lindholm.

\item Thank you to everyone I forgot to mention, including all the unnamed developers of the free, open source software I use in my daily work -- the Open Babel project in particular.

\end{itemize}
Lastly, an even bigger thanks goes to my IRL family and friends, whom I have been seeing much less than I should since I undertook my PhD studies. Thanks, everyone!
\clearpage
\qquad
\vspace{15cm}
\subsubsection*{Licensing}
This work is published under the terms of the Creative Commons Attribution 4.0 International (CC-BY 4.0) license. See \url{http://creativecommons.org/licenses/by/4.0/} for the complete list of license terms.
This work, and all figures and scripts to compile them is available from \url{https://github.com/andersx/phd-thesis/}.
\begin{figure}[h!]
\centering\includegraphics[width=0.4\textwidth]{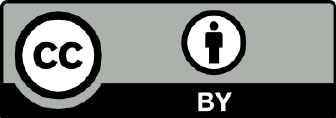}
\end{figure}

\chapter*{Dansk Resumé}
\addcontentsline{toc}{chapter}{Dansk Resumé (Danish Summary)}

Kemien af et protein er tæt forbundet med dens tre-dimensionelle struktur. Af denne grund, er proteinstruktur bestemmelse grundlaget for rationel forståelse af kemien af biologiske processer, der involverer proteiner.

For tiden er flest kendte proteinstrukturer blevet løst ved røntgenkrystallografi. Kravet til løsning af en struktur på denne måde er, at proteinet krystalliserer. Moderne krystaliserings-metoder dog kun har en succesrate på 5\% \cite{xray}. I disse tilfælde kan kernemagnetisk resonans (NMR) metoder anvendes med en vis succes.
I øjeblikket indeholder Protein Data Bank 90.000 strukturer løst ved røntgen- og 9.000 strukturer løst ved NMR-metoder, og omkring 10.000 røntgen- og 500 NMR-strukturer bliver indsendt hvert år \cite{PDB}.

Konventionelle NMR-metoder til bestemmelse af protein strukturer optager et flerdimensionelt spektrum, som korrelerer resonansfrekvenser flere kerner på samme tid.
Fra dette spektrum er først problem at tilordne de kemiske skift af hver kerne. Denne proces er i vid udstrækning automatiseret for hovedkædeatomer, men er mere involveret for sidekædeatomer.  Disse  oplysninger bruges til at identificere toppe i spektret, der svarer til afstandsbegrænsninger (NOE begrænsninger) mellem par af atomer. Disse distance begrænsninger er det bruges til at generere ensembler af strukturer, der tilfredsstiller det givne sæt af begrænsninger.
Protein NMR-spektroskopi har imidlertid flere begrænsninger. Store proteiner har meget overfyldte spektre , hvilket komplicerer opgaven - hovedsagelig på grund af brede toppe og resulterende spektraloverlapning.
Dette er en væsentlig hindring for tilordningen af de kemiske skift og dermed for at finde de
værdifuld NOE begrænsninger. Følgeligt har omkring 95 \% af alle NMR- strukturer i PDB-databasen således har en størrelse på kun 200 aminosyrer eller mindre. Dette kan sammenlignes med de gennemsnitlige størrelser af proteiner i mennesker og \textit{E. coli}, som er henholdsvis omkring 400-600 og 200-400. Problemet kan mindskes ved deuterering  som imidlertid falder til nummer NOE-begrænsninger, der kan findes. Isotopmærkningsmetoder som selektivt mærker visse sidekæder er blevet udviklet som en effektiv strategi for sådanne problemer.

\section*{Computerberegningsmetoder}
En anden tilgang til at løse en proteinstruktur fra aminosyresekvensen er simulering
af energilandskabet af proteinet. Dette kaldes også proteinfoldning. I denne
tilgang, er de mulige konformationer samplet og scoret med en beskrivelse af proteinernes fysik, uden ekstra viden fra eksperimenter. Sådanne \textit{ab initio} tilgange har været anvendet til at bestemme strukturer, typisk med en præcision ned til 3 \AA, via Monte Carlo simuleringer i ROSETTA-programmet \cite{rosetta}. Et andet næveværdigt eksempel er den samtidige bestemmelse af struktur
og dynamik flere små proteiner via meget lange molekylær dynamik (MD) simuleringer med
Anton computer \cite{rdcensemble}.

Selv om disse metoder ikke kræver noget eksperimentelt arbejde, er det ekstremt krævende i
forhold til de edb-ressourcer, der er nødvendige. Desuden er de normalt ikke nemme at konvergere for systemer $>100$ aminosyrer \cite{Lange2012}.
ROSETTA-metoden er (i øjeblikket) velsagtens den mest succesfulde metode til at bestemme
en proteinstruktur via computer beregninger. For nylig viste Baker gruppen, at optagelsen af hovedkæde kemiske skift og RDC data forbedrer ROSETTA-protokollen og tillader bestemmelse af strukturer op til 150 rester \cite{Baker2010,Lange2012}. 

Grundlaget for ROSETTA er fragment-samling af lokale proteinstrukturmodeller, kombineret med raffinering ved hjælp af en energifunktion, der er blevet påvist at fungere bemærkelsesværdigt godt. 
Kort beskrevet  består fuldatom-ROSETTA-energifunktion af flere additive temer som Lennard-Jones potentialer, termer for eksponering solvent, hydrogenbindinger, elektrostatiske par-interaktioner og dispersion-iteraktioner, og endelig torsions potentialer for hovedkæde- og sidekædevinkler. 

Nøjagtigheden af energifunktionen kommer dog på bekostning af beregningsmæssige hastighed og ufuldstændig i den konformationelle prøvetagning, som synes at være den uoverkommelige forhinding for yderligere succes for ROSETTA.
Denne protokol er for nylig blevet forbedret yderligere med inddragelse af meget sparsomme mængder NOE-data \cite{LangePNAS2012}.

Dette gav 7 strukturer omkring 200 aminosyrer, der blev bestemt med en nøjagtighed på mellem 2,5 og 3,9 \AA~fra de tilsvarende eksperimentelle røntgen-strukturer, og desuden blev en god struktur for det 376 aminosyrer store maltosebindingsprotein endda fundet, men dette krævede væsentligt flere NOE oplysninger. Disse simuleringer krævede en 512-kerner supercomputer som kørte i flere dage, for hvert protein.

Et andet nævneværdigt eksempel på protein strukturbestemmelse metoder, der beskæftiger NMR-data, er er CHESHIRE-metoden \cite{cheshire}. CHESHIRE-metoden var den første metode som løste strukturer kun ved brug af kemiske skift, og bruger en fragmentsamlingstilgang, efterfulgt af en Monte Carlo raffinering ved hjælp af et all-atom kraft-felt og en energi-funktion, der inkluderer kemiske skift. Denne metode blev anvendt til at bestemme proteinstrukturer fra kemiske skift, og fandt strukturer for 11 proteiner mellem 54 og 123 aminosyrer i størrelse, til en nøjagtighed på omkring 1,5 \AA~fra de tilsvarende eksperimentelle røntgen-strukturer.
\\\\I det følgende afsnit, er PHAISTOS-programmet introduceret, og formalisme for inkludering af kemiske skift i simuleringer i PHAISTOS er udledt. Dette er et forsøg på at løse
to centrale udfordringer i proteinfoldning: (1) Fuldstændig konformationel prøveudtagning og (2) nøjagtig energi-scoring af konformationelle prøver.
Disse udfordringer er mødt som følger: (1) ved hjælp af en nyudviklet forudindtaget konformationel prøveudtagningsmetode og (2) ved at parametrisere en nøjagtig kemisk skift forudsigelsesmetode, brut med en energifunktion baseret på Bayesiansk statistik, som tillader, at dette kombineres med eksisterende energifunktioner i PHAISTOS.
Denne kombinerede fremgangsmåde vil blive demonstreret på foldningssimuleringer på et testsæt af proteiner med kendte strukturer spænder fra 55 til 269 rester.

\chapter*{Publication list}
\addcontentsline{toc}{chapter}{Publication list}

\section*{List of publications:}
\begin{enumerate}
    \item \underline{Anders S. Christensen}, Stephan P. A. Sauer, Jan H. Jensen (2011) Definitive benchmark study of ring current effects on amide proton chemical shifts. \textit{Journal of Chemical Theory and Computation}, 7:2078-2084.
    \item Wouter Boomsma, Jes Frellsen, Tim Harder, Sandro Bottaro, Kristoffer E. Johansson, Pengfei Tian, Kasper Stovgaard, Christian Andreetta, Simon Olsson, Jan B. Valentin, Lubomir D. Antonov, \underline{Anders S. Christensen}, Mikael Borg, Jan H. Jensen, Kresten Lindorff-Larsen, Jesper Ferkinghoff-Borg, Thomas Hamelryck (2013) PHAISTOS: A framework for Markov chain Monte Carlo simulation and inference of protein structure. \textit{Journal of Computational Chemistry}, 34:1697-1705.
    \item  \underline{Anders S. Christensen}, Troels E. Linnet, Mikael Borg, Wouter Boomsma, Kresten Lindorff-Larsen, Thomas Hamelryck, Jan H. Jensen (2013)  Protein Structure Validation and Refinement Using Amide Proton Chemical Shifts Derived from Quantum Mechanics. \textit{PLoS ONE} 8:e84123.
    \item \underline{Anders S. Christensen}, Thomas Hamelryck, Jan H. Jensen (2014) FragBuilder: An efficient Python library to setup quantum chemistry calculations on peptides models. \textit{PeerJ} 2:e277.
\end{enumerate}
\section*{List of public code:}
\begin{enumerate}
    \item FragBuilder (BSD license) \url{https://github.com/jensengroup/fragbuilder/}
    \item CamShift module (BSD license) \url{https://github.com/jensengroup/camshift-phaistos/}
    \item ProCS module (BSD license) \url{https://github.com/jensengroup/procs-phaistos/}
    \item PHAISTOS (GPL license) \url{http://sourceforge.net/projects/phaistos/}
    \item GAMESS patch FMO-RHF:MP2 (GAMESS license/free) \url{https://github.com/andersx/fmo-rhf-mp2/}
    \item PHAISTOS GUI (BSD license) \url{https://github.com/andersx/guistos/}
    \item NOE module (BSD license) \url{https://github.com/andersx/noe-way-jose/}
\end{enumerate}
\clearpage
\section*{List of other publications:}
\begin{enumerate}
    \item Casper Steinmann, Kristoffer L. Blædel, \underline{Anders S. Christensen}, Jan H. Jensen (2013) Interface of the polarizable continuum model of solvation with semi-empirical methods in the GAMESS program. \textit{PLoS ONE} 8:e67725.
    \item \underline{Anders S. Christensen}, Casper Steinmann, Dmitri G. Fedorov, Jan H. Jensen (2013) Hybrid RHF/MP2 geometry optimizations with the Effective Fragment Molecular Orbital Method. \textit{PLoS ONE} 9:e88800
    \item Jimmy C. Kromann, \underline{Anders S. Christensen}, Casper Steinmann, Martin Korth, Jan H. Jensen (2014) A third-generation dispersion and third-generation hydrogen bonding corrected PM6 method: PM6-D3H+. \textit{PeerJ PrePrints} 2:e353v1.
\end{enumerate}

\thispagestyle{empty}
\renewcommand{\chaptermark}[1]{\markboth{\MakeUppercase{\chaptername\ \thechapter.\ #1}}{}}
\setcounter{page}{1}
\pagenumbering{arabic}
\cleardoublepage
\tableofcontents
\cleardoublepage
\chapter{Introduction}

The chemistry of a protein is tightly linked to its 3-dimensional structure.
For this reason, protein structure determination is the basis of rational understanding of the chemistry of biological processes involving proteins.

Most currently known protein structures have been solved by X-ray crystallography.
One requirement for solving a structure this way is that the protein will crystallize.
Modern crystallization methods, however, only have a success rate of 5\% \cite{xray}.
In these cases, nucleic magnetic resonance (NMR) methods may be used with some success.
Currently the Protein Data Bank contain 90,000 structures solved by X-ray and 9,000 structures solved by NMR methods, and around ~10,000 X-ray and ~500 NMR structures are being submitted each year \cite{PDB}.

Conventional NMR protein structure determination methods records a multidimensional spectrum that correlate the resonance frequencies of several nuclei at the same time.
From this spectrum, the common work flow is to first assign the chemical shifts of each nuclei.
This process is largely automated for backbone nuclei, but is more involved for side chain atoms.
This assignment information is used to identify peaks in the spectrum which correspond to distance restraints (NOE restraints) between pairs of atoms.
These distance restraints are the used to generate ensembles of structures that satisfy the given set of restraints.

Protein NMR spectroscopy, however, has several limitations.
Large proteins have very crowded spectra, which complicates assignment, mostly due to broad peaks and resulting spectral overlap.
This is a substantial hindrance to assignment of the chemical shifts, and therefore obtaining the valuable NOE restraints.
Consequently, around 95\% of all NMR structures in the PDB database thus have a size of only 200 amino acids or less.
This can be compared to the average sizes of proteins in humans and \textit{E. coli}, which are around 400-600 and 200-400, respectively.
These problem can be somewhat alleviated by deuteration which, however, decreases to number NOE restraints that can be obtained.
Isotope labeling schemes which selectively label only certain side chains have been invented, as an efficient strategy for such problems.

\section{Computational methods}
A different approach to solving a protein structure from the amino acid sequence is simulation of the energy landscape of the protein.
This practice is also referred to as \textit{protein folding}.
In this approach, the possible conformations are sampled and scored using a description of the physics of the proteins, with no extra knowledge from experiments.
Such \textit{ab initio} approaches have been used to determine structures up to an accuracy of typically 3 \AA~using Monte Carlo simulations the ROSETTA program \cite{rosetta}.
Another notable example is the simultaneous determination of structure and dynamics of several small proteins via very long molecular dynamics (MD) simulations using the Anton computer \cite{rdcensemble}.

While these methods do not require any experimental input, the are extremely demanding in terms of the computational resources that are required.
Furthermore, they usually fail to converge for structures $>100$ amino acids \cite{Lange2012}.

The ROSETTA methodology is (currently) arguably the most successful method to determine a protein structure computationally.
Recently, the Baker group showed, that inclusion of backbone chemical shifts and RDC data vastly improved the ROSETTA protocol and allowed structures up to 150 residues to be determined \cite{Baker2010,Lange2012}.
The basis of ROSETTA fragment-assembly of local protein structure, combined with  refinement using an energy function that has been demonstrated to work remarkably well.
Briefly described, the all-atom ROSETTA energy function consists of several additive terms such as Lennard-Jones potentials, terms for solvent exposure, hydrogen bonding, electrostatic pair-interactions and dispersion interactions, and finally torsional potentials for backbone and side chain angles.
The demonstrated accuracy of the energy function does come at the cost of computational speed and incomplete conformational sampling seems to be the prohibitive for further success for ROSETTA.
This protocol has recently been further improved with inclusion of very sparse NOE data \cite{LangePNAS2012}.

This allowed 7 structures around 200 amino acids to be determined, to an accuracy of between 2.5 and 3.9 \AA~from the corresponding experimental X-ray structures
Furthermore, a good structure for the 376 amino acids maltose binding protein could even be determined, but this required substantially more NOE data.
These simulations, however required a 512-cores super computer for running several days, for each protein.

Another notable example of protein structure determination methods that employ NMR data is is the CHESHIRE method \cite{cheshire}.
The CHESHIRE method was the first method which solved structures using only chemical shifts, and uses a fragment-assembly approach followed by a Monte Carlo refinement using an all-atom force-field and an energy function that includes chemical shifts.
This method was used to determine the protein structures from chemical shifts, and was demonstrated on 11 proteins between 54 and 123 amino acids in size to an accuracy of around 1.5 \AA~from the corresponding experimental X-ray structures.
\\\\
In the following section, the PHAISTOS program is introduced, and the formalism for inclusion of chemical shifts in simulations in PHAISTOS is derived.
This is an attempt to address the two central challenges in protein folding: (1) complete conformational sampling and (2) accurate energy scoring of conformational samples.

These challenges are met as follows: (1) using a recently developed biased conformational sampling method and (2) by parametrizing an accurate chemical shift predictor and deriving an energy function based rigorously on Bayesian statistics, which allows this to be combined with existing energy functions in PHAISTOS.

The combined approach will be demonstrated on folding simulations on a test-set of protein with known structures ranging from 55 to 269 residues.

\chapter{Introduction to PHAISTOS}

This section servers as an introduction to the PHAISTOS program, and a (very) brief introduction to the theory behind PHAISTOS \cite{Phaistos}. 
This will give the relevant background to read the next chapters.
PHAISTOS is also published and discussed in detail in paper \#2 in this the appendix.

\section{Markov Chain Monte Carlo}

One of the primary goals of simulations in PHAISTOS is to construct the Boltzmann distribution of a protein via Markov chain Monte Carlo (MCMC) sampling for a given potential energy surface at a given temperature.
The Boltzmann distribution of a protein structure, $\mathbf X$, at a given temperature, $T$, is given by:
\begin{equation}
    \label{eq:boltzmann}
    p(\mathbf X) = \frac{1}{Z(T)} \exp{\left( \frac{-E}{k_\mathrm{B}T}\right)},
\end{equation}
where $k_\mathrm{B}T$ is Boltzmann's constant and $Z(T)$ is the partition function at the given temperature.

In Markov chain Monte Carlo the target distribution obtained by repeatedly proposing updates to the current state, and accepting or rejecting these updates with a certain acceptance probability.

It can be shown, that for an infinitely sampled distribution to converge to the correct target distribution, i.e.~$p_\infty(\mathbf X) = p(\mathbf X)$, the Monte Carlo moves that are used to propose updates must satisfy the principle of detailed balance.
That is, the transition from the current state $\mathbf{X}$ to the proposed new state $\mathbf{X'}$ fulfills:
\begin{equation}
    \label{eq:detailed_balance}
    p(\mathbf{X}) p(\mathbf{X} \rightarrow \mathbf{X'}) = 
    p(\mathbf{X'}) p(\mathbf{X'} \rightarrow \mathbf{X})
\end{equation}
where $p(\mathbf{X} \rightarrow \mathbf{X'})$ is the probability to of moving from the state $\mathbf{X}$ to $\mathbf{X'}$ using a given move.
If we further factorize $p(\mathbf{X} \rightarrow \mathbf{X'})$ into an acceptance probability $p_a$ and a move transition probability $p_m$, Eqn. \ref{eq:detailed_balance} gives:
\begin{equation}
    \label{eq:unbiased_mc}
    \frac{p_a(\mathbf{X} \rightarrow \mathbf{X'})}
         {p_a(\mathbf{X'} \rightarrow \mathbf{X})} =
    \frac{p(\mathbf{X'})}
         {p(\mathbf{X})}
    \frac{p_m(\mathbf{X'} \rightarrow \mathbf{X})}
         {p_m(\mathbf{X} \rightarrow \mathbf{X'})}
\end{equation}
Most of the moves in PHAISTOS are symmetric, that is the move bias ratio $p_m(\mathbf{X'} \rightarrow \mathbf{X}) / p_m(\mathbf{X} \rightarrow \mathbf{X'}) = 1$, but for some moves this is not true. 
These biased moves can be exploited to vastly speed up convergence or bias the simulation, and are discussed later in Section \ref{chap:generative}.

\subsection{Metropolis-Hastings}
The simplest Monte Carlo method that satisfies Eqn.~\ref{eq:unbiased_mc} is the Metropolis-Hastings method.
Here a transition $\mathbf{X} \rightarrow \mathbf{X'}$ is accepted using the Metropolis-Hastings acceptance criterion:
\begin{equation}
    \label{eq:mc_mh}
    p_a(\mathbf{X} \rightarrow \mathbf{X'}) = \min \left( 1,
    \frac{p(\mathbf{X'})}
         {p(\mathbf{X})}
    \frac{p_m(\mathbf{X'} \rightarrow \mathbf{X})}
         {p_m(\mathbf{X} \rightarrow \mathbf{X'})} \right)
\end{equation}
Evaluation of the partition function is thus not necessary.
The Metropolis-Hastings method is efficient when exploring native states, and simulations near the critical temperature.
Unfortunately the Metropolis-Hastings method, compared to other MC methods, often gets stuck in local minima, and is therefore generally inefficient when simulating protein folding from an extended strand.

\subsection{Generalized Ensembles}
To avoid the slow convergence problem advanced MC methods are available in PHAISTOS, which emphasize sampling at low energies, which is generally of higher interest in protein structure determination.
These "generalized ensemble" methods are very similar to the Metropolis-Hastings method, and the main difference in the acceptance criterion is that the target distribution $p(\mathbf{X})$ has been replace by a generalized weight function $w(\mathbf{X})$. 
The acceptance criterion then becomes:
\begin{equation}
    \label{eq:mc_gh}
    p_a(\mathbf{X} \rightarrow \mathbf{X'}) = \min \left( 1,
    \frac{w(\mathbf{X'})}
         {w(\mathbf{X})}
    \frac{p_m(\mathbf{X'} \rightarrow \mathbf{X})}
         {p_m(\mathbf{X} \rightarrow \mathbf{X'})} \right)
\end{equation}
Through reweighting, samples from a converged simulation in a generalized ensemble can be reweighted to correspond to the Boltzmann distribution at a given temperature.

PHAISTOS offer two generalized ensemble methods.
In the multicanonical ensemble method, the weight function is $w_\mathrm{muca}(\mathbf{X}) = 1/g(E(\mathbf{X}))$, where $E(\mathbf{X})$ is the energy of the structure $\mathbf{X}$ and $g$ is the associated density of states.
In the inverse-$k$ ensemble, the weight function is given by $w_\textit{1/k}(\mathbf{X}) = 1/k(E(\mathbf{X}))$ where $k(E(\mathbf{X})) = \int_{-\infty}^{E(\mathbf{X})} g(E') dE'$.
The since the density of states is generally unknown, the weight-function is estimated during the simulation.
PHAISTOS uses the MUNINN library to collect histograms of the energy and efficiently provide an estimate of $w(\mathbf{X})$ on-the-fly \cite{muninn}.
\clearpage

\section{Monte Carlo Moves Using Generative Probabilistic Models}
\label{chap:generative}

\begin{figure}
    \centering
    \includegraphics[width=0.75\textwidth]{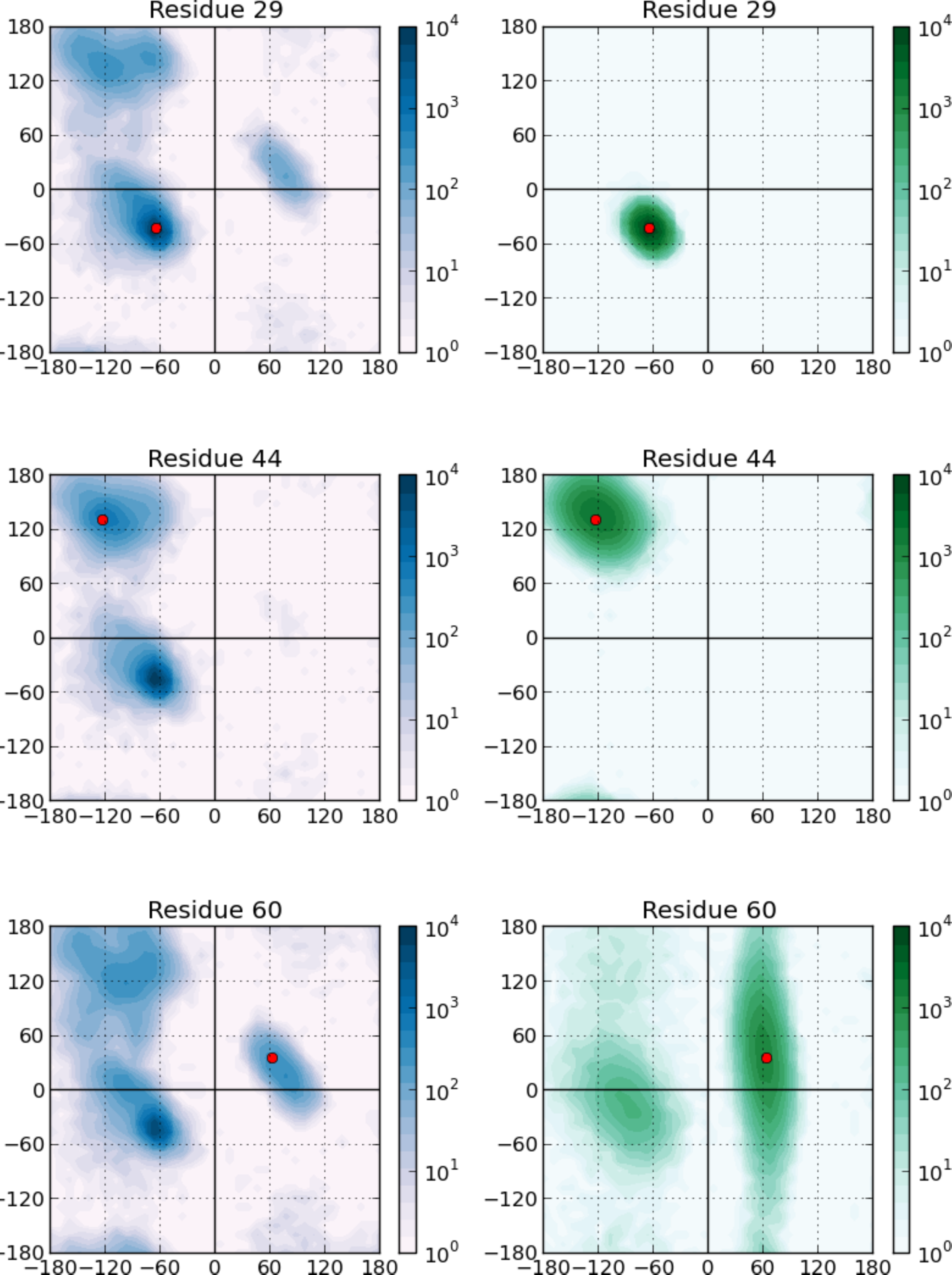}
    \caption{Sampling densities from TorusDBN (left/blue) and TorusDBN-CS (right/green) for the residues 29, 44 and 60 in Ubiquitin. Values from the experimental structure 1UBQ are marked with a red dot. Residue 29 (lysine) is located in the middle of an alpha-helix. Residue 44 (isoleucine) is located in a beta-sheet motif, and finally residue 60 (aspargine) is located in a loop region.}
    \label{fig:torus}
\end{figure}

PHAISTOS proposes new structure samples using a weighted set of difference MC moves, which each randomly changes the current protein structure in a certain way. Briefly, these are divided in side chain moves and backbone moves.
Side chain moves update the rotamer-conformation of a amino-acid single side chain by rotating the dihedral angles on the side chain.
Backbone moves either perform a local perturbation to a strand of a only a few amino acids, or rotates one dihedral angle on the backbone.
\\\\Using random moves which re-sample angles from a uniform distribution, and then constructing a target distribution via an acceptance criterion is a perfectly valid strategy.
However, sampling from a uniform distribution usually lead to slow convergence.
A common approach to alleviate this problem is using fragment assembly, in which small fragments of peptides are assembled from a library of common fragment motifs, such as beta-strands, helices and loops.
This approach, however, introduces a move bias, which must be divided out if the simulation has to obey detailed balance.
Furthermore, it is not clear, how to evaluate the move bias ratio $p_m(\mathbf{X'} \rightarrow \mathbf{X}) / p_m(\mathbf{X} \rightarrow \mathbf{X'})$ when sampling from a fragment library.

A related approach to obtain a similar speed up is biased sampling.
PHAISTOS supports sampling of both side chain and backbone angles from such generative probabilistic models.
In this approach, angles are sampled from distributions that are conditioned on prior knowledge.
Two all-atom generative probabilistic models are supported in PHAISTOS.
TorusDBN which is a hidden-Markov model of backbone angles \cite{Torus08}, and BASILISK \cite{BASILISK} which is a similar model of side chain rotamer-conformations.
Both work are continuous models in torsion-angle space.
The model that is used in this work is TorusDBN, which is is a model that samples backbone dihedral angles conditioned on the amino acid sequence from a distribution that resembles the Ramachandran-plot.
This effectively speeds up convergence of sampling, since uninteresting parts of conformational space in only sampled very rarely.
The importance of the TorusDBN model is discussed in chapter \ref{chapter:results}.

Using models such as TorusDBN and BASILISK introduces a move bias, which compensated for in Eqn.~\ref{eq:unbiased_mc} by multiplying by the ratio $p_m(\mathbf{X'}     \rightarrow \mathbf{X}) / p_m(\mathbf{X} \rightarrow \mathbf{X'})$.
It is possible to determine this ratio, because the likelihood of sampled values can be calculated in the TorusDBN model.
It is thus possible to recover the target distribution (e.g.~the Boltzmann distribution or a generalized ensemble), despite using only biased moves.

Effectively, this turns the target distribution into an effective target distribution.
For sampling from the Boltzmann distribution (e.g.~using a molecular mechanics force field), the effective target distribution becomes
\begin{equation}
    p_e(\mathbf{X}) = p(\mathbf{X}) p_{m}(\mathbf{X}|I),
\end{equation}
where $p_{m}(\mathbf{X}|I)$ is the probability distribution from the generative model, conditioned on the prior information $I$ available to the model.
This is approach is formally equivalent to adding the term $\ln{(p_{m}(\mathbf{X}|I))}$ to the physical energy (although this term does not scale with the temperature):
\begin{eqnarray}
    p_e(\mathbf{X})
    &=& p(\mathbf{X}) p_{m}(\mathbf{X}|I) \nonumber\\
    &\propto& \exp{\left( \frac{-E(\mathbf{X})}{k_{\mathrm{B}}T} \right)}p_{m}(\mathbf{X}|I) \nonumber\\
    &\propto& \exp{\left( \frac{-E(\mathbf{X})}{k_{\mathrm{B}}T} - \ln{(p_{m}(\mathbf{X    }|I))}\right)}
\end{eqnarray}
In other words, biased sampling can be regarded as simply use of a better force field, while the convergence of the simulation is vastly improved.

TorusDBN is implemented in two versions;
standard TorusDBN which, in brief, is conditioned on only the amino-acid sequence, and TorusDBN-CS which is furthermore based on backbone and beta-carbon chemical shifts.
The default TorusDBN model is trained on a set of 1,447 proteins of 180 different SCOP-fold classifications. 
The default TorusDBN-CS model is trained on 1349 proteins and corresponding chemical shifts from the RefDB training set.

Effectively, proposing structures from TorusDBN biases the simulation towards likely angles within the Ramachandran-plot, and furthermore also towards a certain secondary structure type that is likely for the particular amino acid sequence.
The effect of TorusDBN-CS is similar, but the effect is much more pronounced.

Fig.~\ref{fig:torus} shows an example of three different, but typical cases from Ubiquitin. These are alpha-helix, beta-sheet and loop regions.
Residue 29 (lysine) is in a typical alpha helix and this corresponds to the most often sampled cluster from both TorusDBN and TorusDBN-CS. TorusDBN-CS, however, very precisely locates the center of the cluster to within around $\pm 15$ degrees. TorusDBN, in contrast, has some sampling density in the regions typical for beta-sheet and left-handed alpha-helices.

For residue 44 (isoleucine) which is in a typical alpha-helix region of Ubiquitin, TorusDBN-CS accurately pinpoints the distribution of samples around the experimental values.
TorusDBN, however, manages to rule out left-handed helices, but has a higher sampling density in the alpha-helix region than the beta-sheet region. 

The last residue in the examples, residue 60 (aspargine), is located in a loop-region with backbone angles that correspond to a left-handed helix.
Both models sample in the correct region, but TorusDBN favor a regular alpha-helix.
While TorusDBN-CS heavily favors the correct region, angles that are usually not favored in the Ramachandran plot are also frequently sampled in this particular case.
This is presumably due to less fold-diversity in the training set, compared to the set used to train TorusDBN. Generally, however, the TorusDBN-CS distribution is more restrictive than standard TorusDBN.
\subsection{Monte Carlo Moves}

PHAISTOS explores the conformational space by applying local Monte Carlo moves to the protein structure.
Moves are divided into backbone and side chain moves.
All moves work by perturbing one or more internal coordinates.
In principle, all internal coordinates are degrees of freedom.
However, since bond angles and bond lengths are not treated explicitly by the PROFASI force field, these are constrained by the MC moves to standard values \cite{engh-huber}.
Effectively, only dihedral angles are degrees of freedom in the simulations presented here.

This constraint can of course be lifted if the force field include appropriate terms to describe bond angles and bond lengths.
For instance this is supported by the OPLS-AA/L force field included in PHAISTOS, which was used in Paper \#3.

\begin{figure}%
    \centering
    \subfloat[Backbone dihedral angles]{
        {\includegraphics[width=0.62\textwidth]{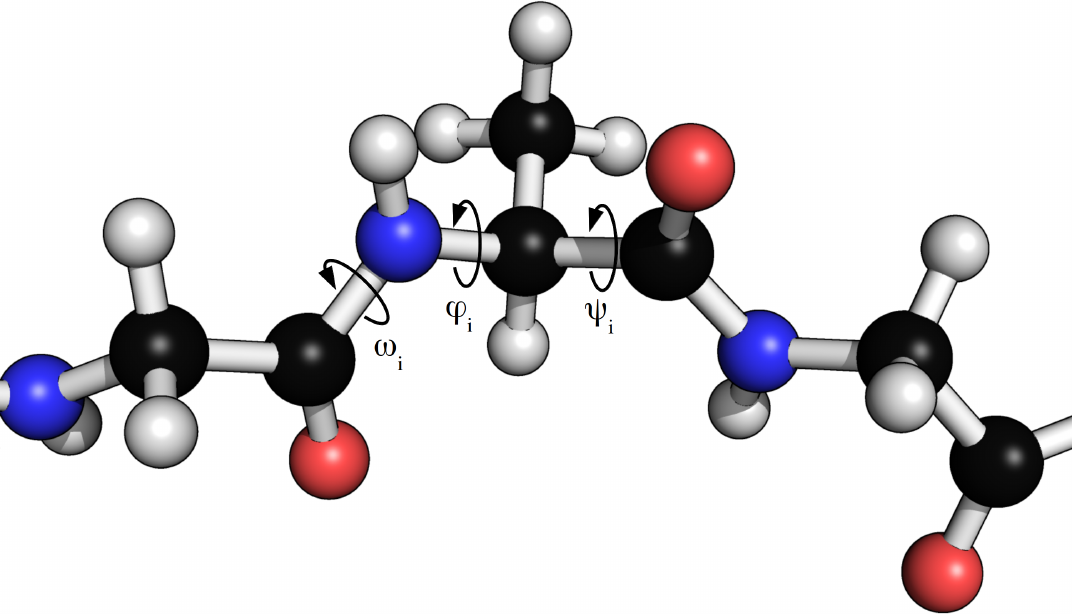} }
    }
    \qquad
    \subfloat[Side chain dihedral angles]{
        {\includegraphics[width=0.28\textwidth]{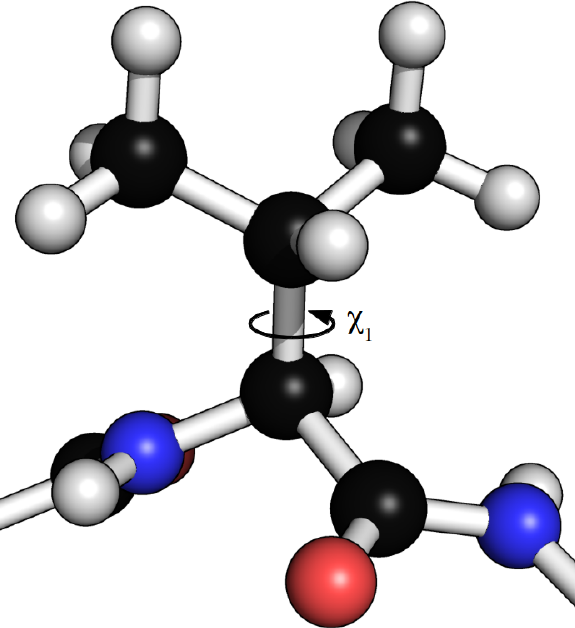} }
    }
    \caption{The degrees of freedoms in a simulations using the PROFASI force field. The $\omega$, $\phi$ and $\psi$ dihedral angles on the backbone are shown in (a) for an alanine residue, and the $\chi_1$ dihedral angle for a valine residue is shown in (b).} 
    \label{fig:dofs}%
\end{figure}

Three different move-types are used in the simulations presented later in this work.
These are introduced below. An overview is displayed in Fig.~\ref{fig:dofs}.
\subsubsection{Pivot Move}
The pivot move re-samples one dihedral angle of the protein backbone. This usually cause large perturbations since two parts of the protein are rotated relative to each other.
As demonstrated later, it is, however, very efficient guiding a folding simulation when biased re-sampling is carried out through TorusDBN or TorusDBN-CS \cite{Torus08}.

\subsubsection{CRISP Move}
In the CRISP move, a number of consecutive residues are selected (default is 7), and the backbone angles of these are perturbed under the constraint that the end-points are fixed in space \cite{crisp}.
This move is particularly efficient at exploring dense states, such as native and near-native states.
This move also supports biased sampling from TorusDBN and TorusDBN-CS.

\subsubsection{Side chain Move}
Side chain moves can either sample new angles uniformly or biased from via BASILISK \cite{BASILISK}.
Additionally, side chain conformations can be drawn from the Dunbrack-rotamer library \cite{dunbrack}.

%
%
%

\chapter{Chemical shifts in a probabilistic framework}

This section introduces the formalism for Monte Carlo simulations which includes both physical energy terms as well as a probabilistic energy terms based on experimentally observed chemical shifts.
The method presented is not new but has not been published in the form presented here.

Working in a probabilistic framework is a powerful strategy for estimation of unknown parameters, and the intention is to present the equations in the form in which they are implemented in PHAISTOS,  so that they can easily be re-implemented in other programs by others.
Simulations using the CamShift and ProCS chemical shifts predictors presented later in this thesis employ the equations presented in this chapter.

\section{Hybrid energy schemes}
There are several ways to include experimental observations in simulations, and combine these with known laws of physics.
A simplistic approach to this problem is to is to define a hybrid energy by defining a penalty function that describes the agreement between experimental data and data calculated from a proposed model with a physical energy (such as from a molecular mechanics force field).
A structure can then be determined, for instance, by minimizing
\begin{equation}
E_{\mathrm{hybrid}}= w_{\mathrm{data}}\ E_{\mathrm{data}}+E_{\mathrm{physical}}.
\label{eq:hybrid_definition}
\end{equation}
where $w_{\mathrm{data}}$ is the weight that quantifies the belief in the energy-model $E_{\mathrm{data}}$ which defines the agreement between the proposed structure and the experimental data relative to the physical energy.

This concept of using a hybrid energy to determine a protein structure was pioneered by Jack and Levitt who simultaneously minimized a molecular mechanics force field energy and the experimental R-factor for the BPTI protein \cite{JackLevitt}.
This approach, however, does not uniquely define neither shape nor weight of $E_{\mathrm{data}}$, and the resulting structure will necessarily depend on these (ill-defined) choices.

Consequently, chemical shifts have been combined with physical energies in a multitude of ways, e.g., weighted RMSD values or harmonic constraints.
The groups of Bax and Baker added the chi-square agreement between SPARTA predicted chemical shift values and experimental chemical shifts with an empirical weight of 0.25 to the ROSETTA all-atom energy \cite{BakerBax}. This methodology was used to determine the structure of 16 small to medium sized proteins.

The CHESHIRE method \cite{cheshire} uses a hybrid energy function, where a classical energy term is divided by the logarithm of a sum of weighted correlation-coefficients between SHIFTX calculated chemical shifts and experimental values.
Here alpha-hydrogen chemical shifts are weighted by a factor of 18 relative nitrogen and carbon chemical shifts which carry a weight of 1.
This hybrid energy is used in the refinement step of the CHESHIRE protocol, and was used to determine the structure of 11 proteins to a backbone RMSD of 1.21 to 1.76 \AA~relative to the corresponding X-ray or NMR structures.

Vendruscolo and co-workers implemented a "square-well soft harmonic potential", and corresponding molecular gradients and were able to run a chemical shift-biased MD simulation using the CamShift chemical shift predictor \cite{CSMD}. Subsequently, the trajectory snapshots were re-weighted by multiplying the chemical shift energy term by an empirical weight of 5.
Using the empirically optimized balance between energy terms, the native state could be determined from the trajectories for 11 small proteins.

In all cases the parameters and weights of $E_{\mathrm{data}}$ had to be carefully tweaked by hand, and it is not clear how to choose optimal parameters.
For instance, different types of chemical shifts may (for optimal results) require different weighting, and a brute-force optimization of all parameters is not straight-forward.

\section{Defining an energy function from Bayes' theorem}
The inferential structure determination (ISD) principles introduced by Rieping, Habeck and Nigles \cite{Rieping2005} defines a Bayesian formulation of Eq.~\ref{eq:hybrid_definition}.
The ISD approach rigorously defines the shape and weight of the $E_{\mathrm{data}}$ term from the definition of an error model, and allows for the weights to be determined automatically as well. In the following section the equations for an ISD approach are derived for combining the knowledge of experimental chemical shifts with a physical energy.

First remember Bayes' theorem which relates a conditional probability (here $A$ given $B$) with its inverse:
\begin{equation}
p\left(A | B \right) =\frac{p\left(B | A \right)p\left(A\right)}{p\left(B \right)}
\end{equation}
Now consider a set of chemical shifts $\{\delta_i\}$, the weight for each chemical shift restraint $\{w_i\}$ in the simulation, and finally the structure to be determined, $\mathbf X$.
This introduces an additional parameter, the weights, which must be determined.
These weights describe the belief in the model that relates a structure to a chemical shift.
In this case, the most likely structure, $\mathbf X$, and optimal choice of $\{w_i\}$ given the set of experimental chemical shifts $\{\delta_i\}$ (via Bayes' theorem) can for instance be found by maximizing:
\begin{eqnarray}
    p\Big(\mathbf X, \{w_i\} \Big| \{\delta_i\} \Big) &=& \frac{p\Big( \{\delta_i\} \Big| \mathbf X, \{w_i\}\Big)p\Big(\mathbf X, \{w_i\} \Big)}{p\Big( \{\delta_i\}\Big)}\nonumber\\
 &\propto& p\Big( \{\delta_i\} \Big| \mathbf X, \{w_i\}\Big)p\Big (\mathbf X, \{w_i\} \Big).
\label{eq:bayes_likelihood}
\end{eqnarray}\\
Here, the \textit{marginal distribution} of $p\left( \{\delta_i\}\right)$ merely serves as a normalizing factor, and can be neglected.
The \textit{likelihood} distribution $p\Big( \{\delta_i\} \Big| \mathbf X, \{w_i\}\Big)$ describes the likelihood of the experimental chemical shifts, given a structure, $\mathbf X$, and the weights $\{w_i\}$.
This requires (1) a forward model to calculate chemical shifts from given structure and (2) an error model that relates the degree of belief in the forward model (that is, the weights) to a probability, based on the difference between experimental and calculated values. 
Later in this chapter, Gaussian and Cauchy distributions are discussed as error models.
The forward model here is a chemical shift predictor, e.g.~CamShift, ProCS, etc.
\\\\If we assume conditional independence, the \textit{prior} $p\Big (\mathbf X, \{w_i\} \Big)$ can be separated as
\begin{equation}
p\Big(\mathbf X, \{w_i\} \Big) = p\Big(\mathbf X\Big) p\Big(\{w_i\} \Big).
\end{equation}
The two priors, $p\Big(\mathbf X\Big)$ and $p\Big(\{w_i\} \Big)$, in brief, describe the distribution of \textit{a priori} meaningful structures (i.e.~usually the Boltzmann distribution), and the probability distribution of the weights, respectively.
In the following $p(\mathbf X)$ is simply the Boltzmann distribution, i.e. 
\begin{equation}
p(\mathbf X) = \frac{1}{Z(T)}\exp{\left(-\frac{E(\mathbf X)}{k_\mathrm{B}T}\right)}
\end{equation}\\
where $E(\mathbf X)$ is the (physical) potential energy of the protein structure, most often calculated using a molecular mechanics force field. $k_\mathrm{B}$ is the Boltzmann constant and $T$ is the temperature of interest.
We need not calculate the partition function, $Z(T)$, because the relative energy landscape is invariant under choice of normalization constant.
Note that $p(\mathbf X)$ also can be introduced via conformational sampling from a biased distribution, such as for example TorusDBN or BASILISK (mimicking the Ramachandran plot and side chain rotamer distributions, respectively). This is discussed later in this chapter.

The prior distribution of the weight parameter $p\Big(\{w_i\} \Big)$ is inherently unknown, except that it is some real number.
One such \textit{uninformative prior} could for instance be a flat distribution over the positive real line.
This distribution, however, may be biased towards very large numbers.
A standard method is to use the Jeffreys' prior, which is a generalization of flat priors, and can be used to model such unknown distributions while introducing only minimal bias.
In the one parameter case the Jeffrey's prior is given as
\begin{eqnarray}
    p(\theta) \propto \sqrt{\mathbf{I}(\theta)},
    \label{eq:jeffreys}
\end{eqnarray}
where $\mathbf{I}(\theta)$ is the \textit{Fisher information} defined (in the one parameter case) as
\begin{eqnarray}
    \mathbf{I}(\theta) = \left\langle \left( \frac{\partial}{\partial\theta} \ln p(x|\theta) \right)^2 \right\rangle.
\end{eqnarray}
The corresponding priors for the Gaussian and Cauchy distributions are discussed in the next sections.

\subsection{Gaussian error model}
Selecting an error model is the basic assumption that difference (the error) between a chemical shift calculated from a structure and the corresponding experimentally measured chemical shift, given as $\Delta\delta_i(\mathbf X) = \left| \delta_i^{\mathrm{predicted}}(\mathbf X) - \delta_i^{\mathrm{experimental}}\right|$, is distributed according to some defined distribution.
Following the principle of maximum entropy, the Gaussian distribution is the least biasing distribution, and is the least biasing choice of error model.
In this case, the weight parameter introduced in the previous section corresponds to the standard deviation, $\sigma$ of the Gaussian distribution.
For simplicity, it is assumed that the mean of the Gaussian is zero.
The total likelihood is then the product of the probability of each $\Delta\delta_i(\mathbf X)$:

\begin{eqnarray}
p\Big( \{\delta_i\} \Big| \mathbf X, \{\sigma_i\}\Big) & = & \prod_{i=0}^{n} p\left( \Delta\delta_i (\mathbf X)| \sigma_i \right)\nonumber\\
& \propto & \prod_{i=0}^{n} \frac{1}{\sigma_i } \exp{ \Bigg( - \frac{\Delta\delta_{i}(\mathbf X)^2}{2\sigma_i^2} \Bigg) }
    \label{eq:gauss_lik}
\end{eqnarray}
Next we derive Jeffreys' prior for the uncertainty of a generic Gaussian distribution of the form
\begin{eqnarray}
    p(x|\mu, \sigma) = \frac{1}{\sqrt{2\pi\sigma^2}} \exp \left( \frac{-(x-\mu)}{2\sigma^2} \right).
\end{eqnarray}
Via Eqn.~\ref{eq:jeffreys}, this immediately gives us the Jeffreys' prior:
\begin{eqnarray}
    p(\sigma)
    & \propto & \sqrt{\left\langle \left( \frac{\partial}{\partial\sigma}
        \ln p(x|\mu, \sigma) \right)^2 \right\rangle}\nonumber\\
    & = & \sqrt{\left\langle \left( \frac{\partial}{\partial\sigma}
        \ln \left[\frac{1}{\sqrt{2\pi\sigma^2}} \exp \left( \frac{-(x-\mu)}{2\sigma^2} \right) \right]
        \right)^2 \right\rangle}\nonumber\\
    & = & \sqrt{\left\langle \left(\frac{(x-\mu) - \sigma^2}{\sigma^3} \right)^2 \right\rangle}\nonumber\\
    & = & \sqrt{\int^{\infty}_{-\infty}  p(x|\mu, \sigma) \left(\frac{(x-\mu) - \sigma^2}{\sigma^3} \right)^2 dx}\nonumber\\
    & = &\sqrt{ \frac{2}{\sigma^2}} \ \propto \ \frac{1}{\sigma}
        \label{eq:gauss_prior}
\end{eqnarray}
Practically, it is impossible to have a separate weight for each individual chemical shift, and the chemical shift of nuclei of the same type thus carry the same weight.
The forward model is similar for all nuclei of the same type, so this is somewhat well-justified.
\\\\In the following equations, $j$ runs over atom types (e.g.~C$^\alpha$ or H$^\alpha$, etc), and $i$ over residue number.
Inserting Eqn.~\ref{eq:gauss_lik} and Eqn.~\ref{eq:gauss_prior} into Eqn.~\ref{eq:bayes_likelihood}, we arrive at a total probability of:
\begin{eqnarray}
    p\Big(\mathbf X, \{\sigma_j\} \Big| \{\delta_{ij}\} \Big) 
    &\propto& p\Big( \{\delta_{ij}\} \Big| \mathbf X, \{\sigma_j\}\Big)p\Big (\mathbf X \Big) p\Big ( \{\sigma_j\} \Big) \nonumber\\
    &\propto&   \prod_{j=0}^{m} \prod_{i=0}^{n} \frac{1}{\sigma_j} \exp{ \Bigg( - \frac{\Delta\delta_{ij}(\mathbf X)^    2}{2\sigma_j^2} \Bigg) }
    \exp{\left(-\frac{E(\mathbf X)}{k_\mathrm{B}T}\right)}
    \prod_{j=0}^{m} \frac{1}{\sigma_j}\nonumber\\
    & = & \prod_{j=0}^{m} \left( \frac{1}{\sigma_j }\right)^{n} \exp{ \left( \sum_{i=0}^{n} - \frac{\Delta\delta_{ij}(\mathbf X)^2}{2\sigma_j^2} \right)\exp{\left(-\frac{E(\mathbf X)}{k_\mathrm{B}T}\right)}}  \prod_{j=0}^{m} \frac{1}{\sigma_j}\nonumber\\
    & = & \prod_{j=0}^{m} \left( \frac{1}{\sigma_j }\right)^{n+1} \exp{ \left( \sum_{i=0}^{n} - \frac{\Delta\delta_{ij}(\mathbf X)^2}{2\sigma_j^2} \right)\exp{\left(-\frac{E(\mathbf X)}{k_\mathrm{B}T}\right)}}
    \label{eq:gauss_joint}
\end{eqnarray}\\
This can be converted to the corresponding hybrid-energy:
\begin{eqnarray}
    E_{\mathrm{hybrid}}
    & = &- k_\mathrm{B}T \ln{ \Big(p\Big( \mathbfit X, \{\sigma_i\} \Big| \{\delta_{ij}\} \Big)\Big) }\nonumber\\
    & = & E(\mathbfit X) + k_\mathrm{B}T \sum_{j=0}^{m}(n+1) \ln{ \left(\sigma_j \right)} + k_\mathrm{B}T \sum_{j=0}^{m} \sum_{i=0}^{n} \frac{    \Delta\delta_{ij}(\mathbf X)^2}{2\sigma_j^2}
\end{eqnarray}
This expression, except for the term $(n+1) \ln{ \left(\sigma \right)}$, is essentially an energy function using harmonic constraints.
It is, however, the balance between the two terms which include $\sigma$ that makes things work.
The term $(n+1) \ln{ \left(\sigma \right)}$ yields the lowest energy for small values of $\sigma$, while the term $\frac{\Delta\delta (\mathbf X)^2}{2\sigma^2}$ is lower for large values of $\sigma$.

Furthermore, the effect of the prior is minute: Using Jeffreys' prior this term is $(n+1) \ln{\left(\sigma\right)}$, whereas using a uniform prior the same term is $ n \ln{\left(\sigma\right)}$. Since $n$ is the number of measured chemical shifts of a certain type, the value is usually in the order of $\sim 100$.

\subsection{Cauchy error model}
Due to numerical instabilities in simulation using the Gaussian error model, a similar model was derived, using a Cauchy distribution as error model. The most notable difference between the Gaussian and Cauchy distributions is that the Cauchy distribution has fatter tails, and thus allows for larger outliers. The differences are  discussed in further detail in the Results section in this chapter.
\\\\Similarly to Eqn.~\ref{eq:gauss_lik}, we assume that the location parameter of the Caucy-distribution is zero, and  use the scale-parameter, $\gamma$ as the weight. The total likelihood is then:

\begin{eqnarray}
    p\Big( \{\delta_i\} \Big| \mathbf X, \{\gamma_i\}\Big) 
    & = & \prod_{i=0}^{n} p\left( \Delta\delta_i (\mathbf X)| \gamma_i \right)\nonumber\\
    & \propto & \prod_{i=0}^{n} \frac{1}{\gamma_i \left[ 1+ \left(\frac{\Delta\delta_{i}(\mathbf X)}{\gamma_i}\right)^2\right]}
    \label{eq:cauchy_lik}
\end{eqnarray}
And for the $\gamma$ parameter of the generic Cauchy distribution of the form
\begin{eqnarray}
    p(x|x_0, \gamma) = \frac{1}{\pi\gamma\left[ 1 + \left(\frac{x-x_0}{\gamma} \right)^2\right]},
\end{eqnarray}
we obtain the following Jeffreys' prior:
\begin{eqnarray}
p(\gamma)
& \propto & \sqrt{\left\langle \left( \frac{\partial}{\partial\gamma}
    \ln p(x|x_0, \gamma) \right)^2 \right\rangle}\nonumber\\
& = & \sqrt{\left\langle \left( \frac{\partial}{\partial\gamma}
\ln \left[\frac{1}{\pi\gamma\left[ 1 + \left(\frac{x-x_0}{\gamma} \right)^2\right]} \right] \right)^2 \right\rangle}\nonumber\\
& = & \sqrt{\left\langle \left( -\frac{\gamma^2 - (x-x_0)^2}{\gamma^3 +\gamma(x-x_0)^2} \right)^2 \right\rangle }\nonumber\\
& = & \sqrt{\int^{\infty}_{-\infty}  p(x|x_0, \gamma) \left(  -\frac{\gamma^2 - (x-x_0)^2}{\gamma^3 +\gamma(x-x_0)^2}\right)^2 dx }\nonumber\\
& = & \sqrt{\frac{1}{2\gamma^2}} \ \propto \ \frac{1}{\gamma} 
\label{eq:cauchy_prior}
\end{eqnarray}
Again, it is practically impossible to have a separate weight for each individual chemical shift, and the chemical shift of nuclei of the same type thus carry the same weight.
In the following equations, $j$ runs over atom types (e.g.~C$^\alpha$ or H$^\alpha$, etc), and $i$ over residue number.
Assembling the Eqn.~\ref{eq:cauchy_lik} and Eqn.~\ref{eq:cauchy_prior} into Eqn.~\ref{eq:bayes_likelihood}, we arrive at the total probability of:
\begin{eqnarray}
    p\Big(\mathbf X, \{\gamma_j\} \Big| \{\delta_{ij}\} \Big) 
    &\propto& p\Big( \{\delta_{ij}\} \Big| \mathbf X, \{\gamma_j\}\Big)p\Big (\mathbf X, \{\gamma_j\} \Big) \nonumber\\
    &\propto&   \prod_{j=0}^{m} 
    \prod_{i=0}^{n} \frac{1}{\gamma_j \left[ 1+ \left(\frac{\Delta\delta_{ij}(\mathbf X)}{\gamma_j}\right)^2\right]}
    \exp{\left(-\frac{E(\mathbf X)}{k_\mathrm{B}T}\right)}
    \prod_{j=0}^{m} \frac{1}{\gamma_j}\nonumber\\
    &= &   \prod_{j=0}^{m} \left( \frac{1}{\gamma_j} \right)^{n + 1}
    \prod_{i=0}^{n} \frac{1}{ 1+ \left(\frac{\Delta\delta_{ij}(\mathbf X)}{\gamma_j}\right)^2}
    \exp{\left(-\frac{E(\mathbf X)}{k_\mathrm{B}T}\right)}
\end{eqnarray}\\
The associated hybrid energy is then given as:
\begin{eqnarray}
    E_{\mathrm{hybrid}}
    & = &- k_\mathrm{B}T \ln{ \Big(p\Big( \mathbfit X, \{\gamma_i\} \Big| \{\delta_{ij}\} \Big)\Big) }\nonumber\\
    & = & E(\mathbf X) 
    + k_\mathrm{B}T \sum_{j=0}^{m}(n+1) \ln{ \left(\gamma_j \right)} 
    + k_\mathrm{B}T \sum_{j=0}^{m} \sum_{i=0}^{n} \ln{\left[ 1 + \left(\frac{\Delta\delta_{ij}(\mathbf X)}{\gamma_j} \right)^2\right]}
\end{eqnarray}

\subsection{Marginalization of Weighting parameter}

A third option also explored here, is the removal of the weight parameter by projection.
This procedure is known as \textit{marginalization}, and is carried out by integrating over all values of the weight parameter.
While integration is straight-forward for the Gaussian error-model, the similar expression for the Cauchy distribution does not integrate easily, and the Cauchy-model was not investigated here.
From the joint probability distribution in Eqn.~\ref{eq:gauss_joint} we obtain the following:

\begin{eqnarray}
    p_{\mathrm{marginal}}\Big(\mathbf X \Big| \{\delta_{ij}\} \Big) 
    & = & \int_0^\infty p\Big( \{\delta_{ij}\} \Big| \mathbf X, \{\sigma_j\}\Big)p\Big (\mathbf X \Big) p\Big ( \{\sigma_j\} \Big) d\sigma \nonumber\\
    & = & \int_0^\infty \prod_{j=0}^{m} \left( \frac{1}{\sigma_j }\right)^{n+1} \exp{ \left( \sum_{i=0}^{n} - \frac{\Delta\delta_{ij}(\mathbf X)^2}{2\sigma_j^2} \right)\exp{\left(-\frac{E(\mathbf X)}{k_\mathrm{B}T}\right)}} d\sigma\nonumber\\
    & = &  \prod_{j=0}^{m}  \left( \sum_{i=0}^{n} \Delta\delta_{ij}(\mathbf X)^2 \right)^{n/2} \exp{\left(-\frac{E(\mathbf X)}{k_\mathrm{B}T}\right)}
\end{eqnarray}\\
The hybrid energy associated with the marginalized probability is then given as:
\begin{eqnarray}
    E_{\mathrm{hybrid}}
    & = &- k_\mathrm{B}T \ln{ \Big(p_{\mathrm{marginal}}\Big( \mathbfit X \Big| \{\delta_{ij}\} \Big)\Big) }\nonumber\\
    & = & E(\mathbf X) 
    + \frac{n}{2} \sum_{j=0}^{m} \ln{\sum_{i=0}^{n} \Delta\delta_{ij}(\mathbf X)^2}
\end{eqnarray}

\subsection{Soft Square-Well Energy Function}
The last type of hybrid energy term explored here, is a potential designed specifially for molecular dynamics simulations biased by the CamShift predictor \cite{robustelli2009, CSMD}.
In this case, the hybrid-energy is given as:
\begin{eqnarray}
    E_{\mathrm{hybrid}}
    = E(\mathbf X) + \alpha E_{\mathrm{CS}}(\mathbf X, \{\delta_{ij}\}),
\end{eqnarray}
where $E_{\mathrm{CS}}(\mathbf X, \{\delta_{ij}\})$ is an empirically derived penalty function that has been demonstrated through simulations to work well for protein structure determination.
$\alpha$ is a weight parameter which was set to 1 during simulation.
This penalty function is termed a "soft-square harmonic well", and given by:

\begin{equation}
    E_{\mathrm{CS}}(\mathbf X, \{\delta_{ij}\}) = \sum_{j=0}^{m} \sum_{i=0}^{n} E_{ij},
\end{equation}
with 
\begin{equation}
    E_{ij} = \begin{cases} 
        0
                & \mbox{if }  \Delta\delta_{ij}(\mathbf X) < n\epsilon_j\\ 
        \left( \frac{\Delta\delta_{ij}(\mathbf X) - n \epsilon_j}{\beta_j}\right)^2
                & \mbox{if }  n\epsilon_j < \Delta\delta_{ij}(\mathbf X) < x_0 \\ 
        \left( \frac{x_0- n \epsilon_j}{\beta_j}\right)^2 + \gamma \tanh{
            \frac{2(x_0- n)(\Delta\delta_{ij}(\mathbf X) - x_0)}{\gamma\beta_j^2}
        }
                & \mbox{if }  x_0 \leq \Delta\delta_{ij}(\mathbf X) .
    \end{cases}
\end{equation}
where the parameters, $n\epsilon_j, x_0, \beta_j$ and $\gamma$ have been empirically adjusted.
The potential has a flat bottom, with the width of $n\epsilon_j$.
The flat bottom corresponds to the expected standard deviation of CamShift, to avoid overfitting in the simulation.
The penalty function grows harmonically until a cut-off of $x_0$ and follows a somewhat flat hyperbolic tangent function after this.
While there is no substantial theoretical backing

\section{Sampling strategy for weight parameters}
Since the nuisance parameters of the energy functions are unknown, they too must be sampled.
The move used to update the value of the nuisance parameters must obey detailed balance:
\begin{eqnarray}
    p\left(w \rightarrow w'\right) = p\left(w' \rightarrow w\right)
\end{eqnarray}
The simplest Monte Carlo move is simply adding a number from a normal distribution with $\mu = 0$, this clearly obeys detailed balance, since the distribution is symmetric.
For the weight parameters, $\gamma$ and $\sigma$, of the Cauchy and Gaussian distributions, respectively, we found a variance of $0.05$ in the normal distributed move to converge quickly and stably.

\subsection{Molecular mechanics force field}

One reasonable prior distribution for protein structure, $p(\mathbf{X)}$, is the Boltzmann distribution, e.g.:

\begin{equation}
    p(\mathbf{X}) \propto \exp\left( \frac{-E}{k_\mathrm{B}T}\right)
\end{equation}
where $E$ is the energy of the structure, $\mathbf{X}$ and $k_\mathrm{B}$ and $T$ are Boltzmann's constant and the temperature, respectively.
The energy of the structure is in this context usually approximated by a molecular mechanics force field that is taylor-made for protein simulations. PHAISTOS currently supports two different protein force field: The OPLS-AA/L force field with a GB/SA solvent term, and the coarse-grained PROFASI force field. 
The OPLS-AA/L is an all-atom force field with an additional solvation. The PROFASI force field is a coarse-grained force-field which assumes fixed bond-lengths and angles and furthermore has a very aggressive 4.5 \AA~cut-off of long-range interaction terms.

\section{Results}

\subsection{Results -- sampling of weight parameters}
\begin{figure}%
    \centering
    \subfloat[Gaussian distribution]{
        {\includegraphics[width=0.45\textwidth]{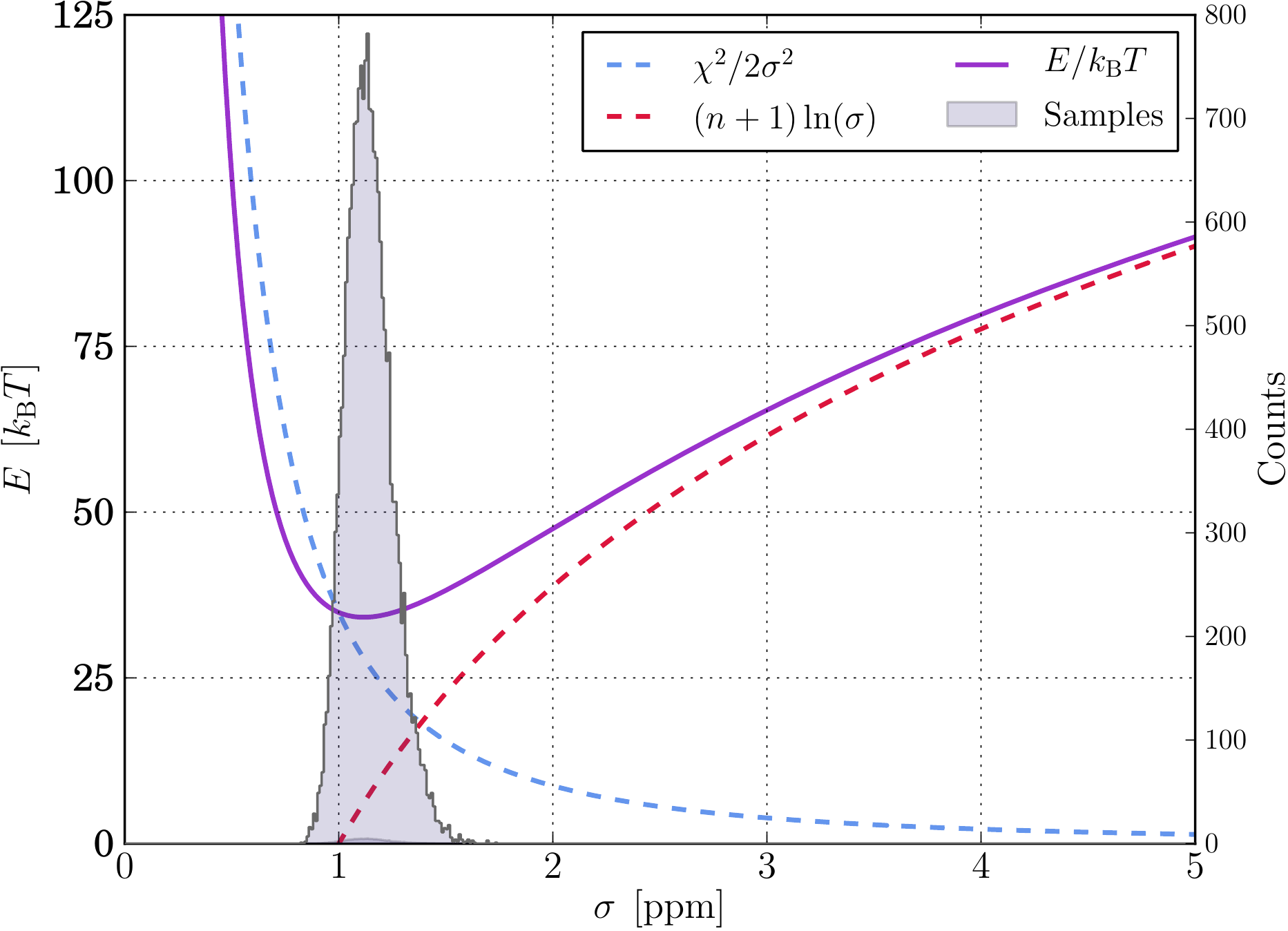} }
    }
    \qquad
    \subfloat[Cauchy distribution]{
        {\includegraphics[width=0.45\textwidth]{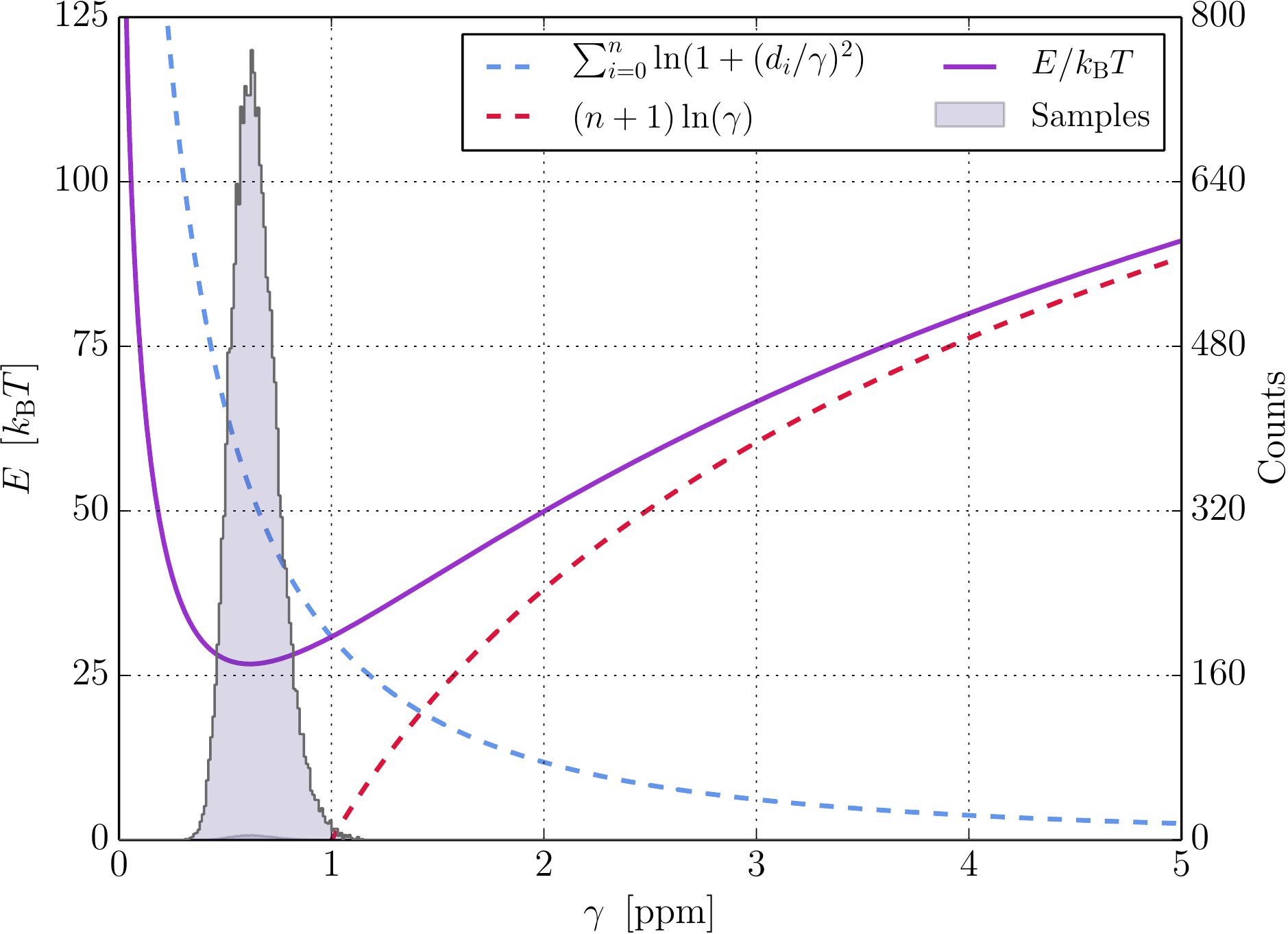} }
    }
    \caption{Sampling of $\sigma$ and $\gamma$ for 2OED for Ca-chemical shifts. In this example $n = 55$ and $\chi^2 = 69.7$. Sampled values of the weight parameters clearly cluster around the minimum of the energy function.}
    \label{fig:example}%
\end{figure}

Figure \ref{fig:example} show a histogram of 100,000 sampled values of $\gamma$ and $\sigma$ for the NMR structure of Protein G (PDB-id: 2OED). 
No structural moves were used, and the results are thus temperature independent since the physical energy is constant.
A total of 55 C$^\alpha$ experimental chemical shifts were used in this example (RefDB-id: 2575), and CamShift was used to calculate the chemical shifts. 
The initial values of $\sigma$ and $\gamma$ was 10.0, in order to demonstrate the stable convergence using the simple move.

In both simulations, the sampling algorithm converges sampling around the minimum of the energy function.
In both cases, these minima are in very good agreement with the values calculated by the test set that was used to validate the performance of CamShift.
The largest sampled bins are centered on $\sigma = 1.26$ ppm and $\gamma = 0.63$ ppm for the Gaussian and Cauchy distributions, respectively.
These number can be compared to the maximum likelihood estimates (MLE) obtained on the 7 protein benchmark set used to determine the accuracy of Camshift.
Here the values are $\sigma = 1.3$ ppm and $\gamma = 0.7$ ppm for the Gaussian and Cauchy distributions, respectively.

\subsection{Performance of energy functions}

Here folding simulation using 11 different variations of the energy function derived and mentioned previously are compared.
All energy functions have been implemented in the CamShift module in PHAISTOS, which was also used to run all simulations.
The test were carried out on Protein G and the engrailed homeodomain (ENHD). The reference structures were the structures 2OED and 1ENH.
An overview of the different simulation types can be found in Table~\ref{tab:prior_folds}.
For each energy function, 20 independent simulations were carried out for a total of 50,000,000 MC steps each.
Each simulation was initialized from a different random, extended strand.
Maximum likelihood estimated (MLE) values of the $\sigma$ and $\gamma$ weight parameters estimated take from the 7-protein test set reported in reference \cite{CamShift}.
For simulations where the weight parameter was sampled, an additional 500,000 Monte Carlo steps were carried out corresponding to the extra moves required to sample this weight (the computational overhead of these 500,000 moves is negligible).
Chemical shifts were calculated using the CamShift module.
All simulations used the PROFASI force field and sampling from either TorusDBN or TorusDBN-CS.
\begin{table}[h]
    \caption{Protocols used in the comparison of energy functions and success rates.}
    \begin{center}
    \begin{threeparttable}
    \begin{tabular}{l l l l l l}
Energy type & Weight  & TorusDBN-mode & Sampling Bias & Correct sampling\tnote{a} & Correct scoring\tnote{b}\\\hline
Gauss         & Fixed/MLE   & Torus       & Biased   & 20/20  & 2/6  \\
Gauss         & Sampled & Torus       & Biased   &  0/0   & 0/0  \\
Cauchy        & Fixed/MLE   & Torus       & Biased   & 20/12  & 7/4  \\
Cauchy        & Sampled & Torus       & Biased   & 20/5   & 6/1  \\
Cauchy        & Sampled & Torus-CS    & Biased   & 20/20  & 4/2  \\
Cauchy        & Sampled & Torus       & No bias  &  0/0   & 0/0  \\
Cauchy        & Sampled & Torus-CS    & No bias  &  0/0   & 0/0  \\
Square-well   & Fixed   & Torus       & Biased   &  1/2   & 1/0  \\
Marginalized  & N/A     & Torus       & Biased   &  7/17  & 1/8  \\
No CS         & N/A     & Torus       & Biased   &  0/0   & 0/0  \\
No CS         & N/A     & Torus-CS    & Biased   &  8/10  & 2/0
    \end{tabular}
    \begin{tablenotes}
        \item[a] Number of threads with a CA-RMSD of $<5$ \AA\ (using all residues). Listed as xx for Protein G and yy for ENHD, i.e.~xx/yy.\\
        \item[b] Number of threads where the lowest energy sample has a CA-RMSD of $<3$ \AA\ (using all residues). Listed as xx for Protein G and yy for ENHD, i.e.~xx/yy.\\
    \end{tablenotes}
    \end{threeparttable}
    \end{center}
    \label{tab:prior_folds}
\end{table}

\begin{figure}%
    \centering
    \includegraphics[width=0.75\textwidth]{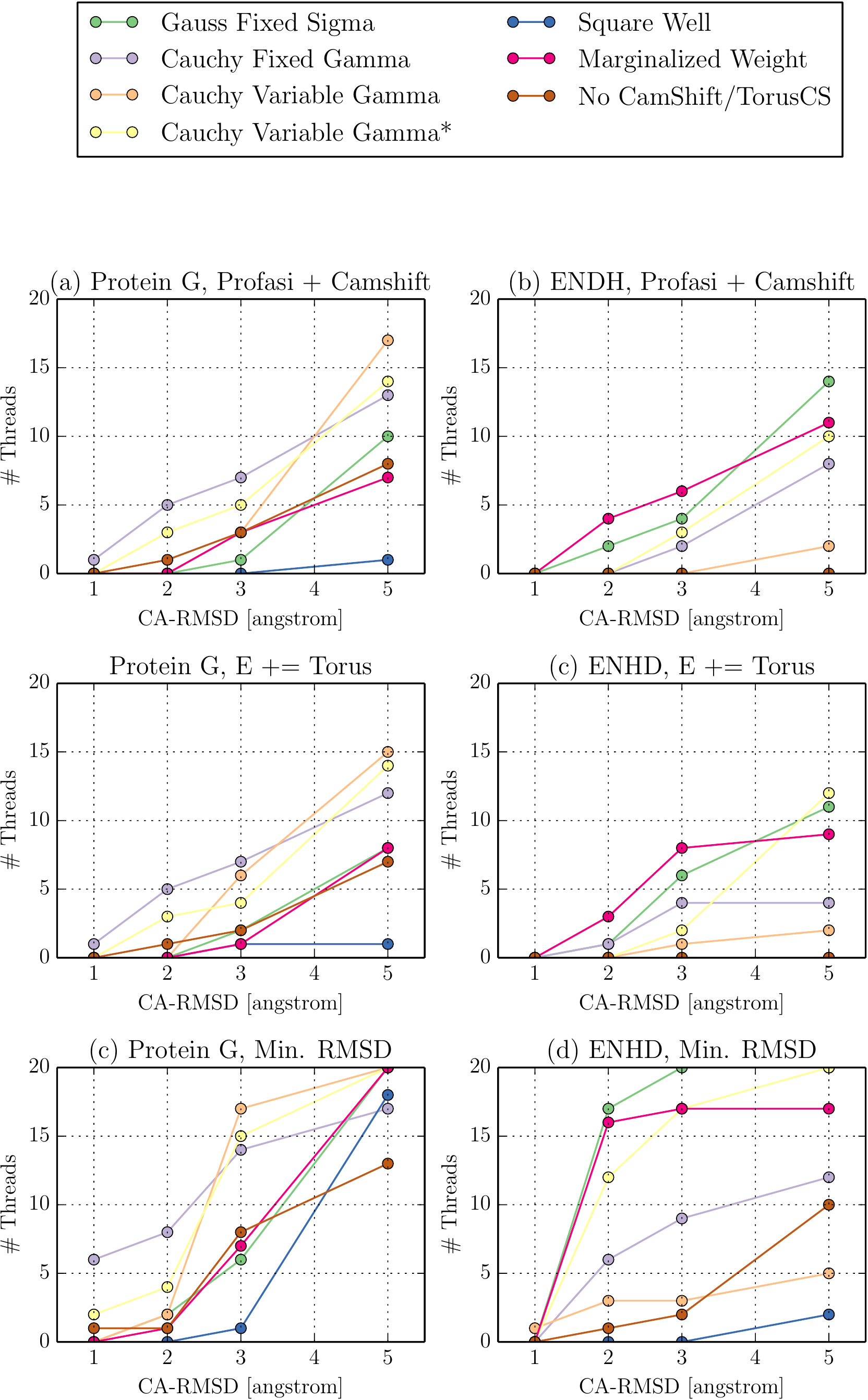}
    \caption{Overview of folding simulations using 7 different chemical shift energy types. Sampling was biased by TorusDBN and the PROFASI energy term was used as well.
             In (a) and (b) the number of threads where the lowest energy samples are under thresholds of 1, 2, 3 and 5 \AA~CA-RMSD from the crystal structure is plotted. 
             The energy here is calculated as the PROFASI energy multiplied by $k_{\mathrm{B}}T$ plus the chemical shift energy term. 
In (c) and (d), the log-likelihood from TorusDBN has been added to the total energy.
In (e) and (f), the number of threads in which samples are found below under thresholds of 1, 2, 3 and 5 \AA~CA-RMSD from the crystal structure is plotted. 
*In this simulation TorusDBN-CS is used instead of TorusDBN.}
    \label{fig:cauchy_results}%
\end{figure}

In two simulations, the bias was removed from the simulation, which corresponds to an unbiased simulation.
Two reference simulations were carried out with no chemical shift energy-function, in order to analyze the effect of sampling from TorusDBN and the effect of the PROFASI force field.
The simulations used a mix of 40\% biased CRISP-moves, 10\% biased pivot moves and 50\% uniform side chain moves.
The simulation was carried out in the multicanonical ensemble via MUNINN.
Minimum and maximum $\beta$-values were set to 0.3 and 1.05, and the temperature was set to 300K.
In all simulations, the number of threads which had samples below thresholds of 5, 3, 2 and 1 \AA~CA-RMSD from the crystal structure was recorded.
Similarly, the number of threads in which the lowest energy structure was below thresholds of 5, 3, 2 and 1 \AA~CA-RMSD from the crystal structure was recorded.
These figures are used to analyze whether sampling or correct energy scoring is are limiting factors in the particular simulations.
The energy was calculated as the PROFASI energy multiplied by $k_{\mathrm{B}}T$ plus the chemical shift energy term plus the log-likelihood calculated from TorusDBN.
An overview of these results can be seen in Fig.~\ref{fig:cauchy_results} (only simulations that had any samples below 5 \AA~CA-RMSD from the crystal structure are shown).

For both proteins, using a Gaussian model and sampling the $\sigma$ uncertainty does not lead to meaningful values for $\sigma$.
In short, PHAISTOS is able to generate a structure which has no difference between experimental and calculated chemical shifts for a certain atom type.
Consequently, the value of $\sigma$ converges to zero, which effectively freezes the structure in the simulation.
The simulations in which the move-bias from TorusDBN and TorusDBN-CS was removed did not sample any structures below 

For simulations using Gaussian or Cauchy types of energy function all thread had samples below 5 \AA~CA-RMSD from the crystal structure for Protein G and between 5-20 for ENHD 
In the simulation using the square-well potential only 1 thread had samples below 5 \AA~for Protein G  and only 2 for ENHD.
For the simulation with marginalized weight parameters, the same figures were 7 and 17, respectively.
The reference simulations with no chemical shift in the energy function had no samples below 5 \AA~for biased sampling from TorusDBN, but 8 and 10 threads below 5 \AA~for biased sampling from TorusDBN-CS.

Comparing the number of threads for which the lowest energy sample was below 3 \AA~CA-RMSD from the crystal structure.
For both proteins, using fixed weights is somewhat better than using sampled weights with the Cauchy distribution.
The result for the square-well potential cannot be interpreted to a statistical significance because only one and two threads were close to the correct fold, but one thread correctly identified the folded state below 3 \AA~CA-RMSD as the lowest energy for Protein G.

In conclusion, the Gauss and Cauchy error models perform well in sampling and scoring.
The fixed MLE weights seem to be work equally well to sampling weights for the cauchy distribution, with no substantial differences.
The performance of the energy with marginalized weights generally performed worse in guiding the sampling, but well in scoring samples for ENHD.
The square-well potential did not improve the sampling much.
The reason why it has previously been shown to work well, might be that it was combined with a better force-field (AMBER03) to which it was specifically designed.
One clear conclusion is that it is useful to not remove the bias from TorusDBN, and keeping the TorusDBN-CS bias seems guide folding significantly more. Even though this formally constitutes is double-counting of effect of knowledge about chemical shifts, this practice seemingly has no adverse effects.

\chapter{Graphical User Interface for PHAISTOS}

Setting up simulations in PHAISTOS requires expert knowledge about the program. 
Firstly, while all modules and settings have reasonable default settings, there are still many things that cannot be specified via default alone, and secondly, the complete list of settings in PHAISTOS is around 2500 options that must be set or taken as default values.

In order to make PHAISTOS more attractive to new users, I wrote a GUI can set up most simulations for most of the simulations covered by this thesis.
The GUI for PHAISTOS is aptly named Guistos and is written in Python 2.x using TkInter.

\begin{figure}
    \centering
    \includegraphics[width=0.70\textwidth]{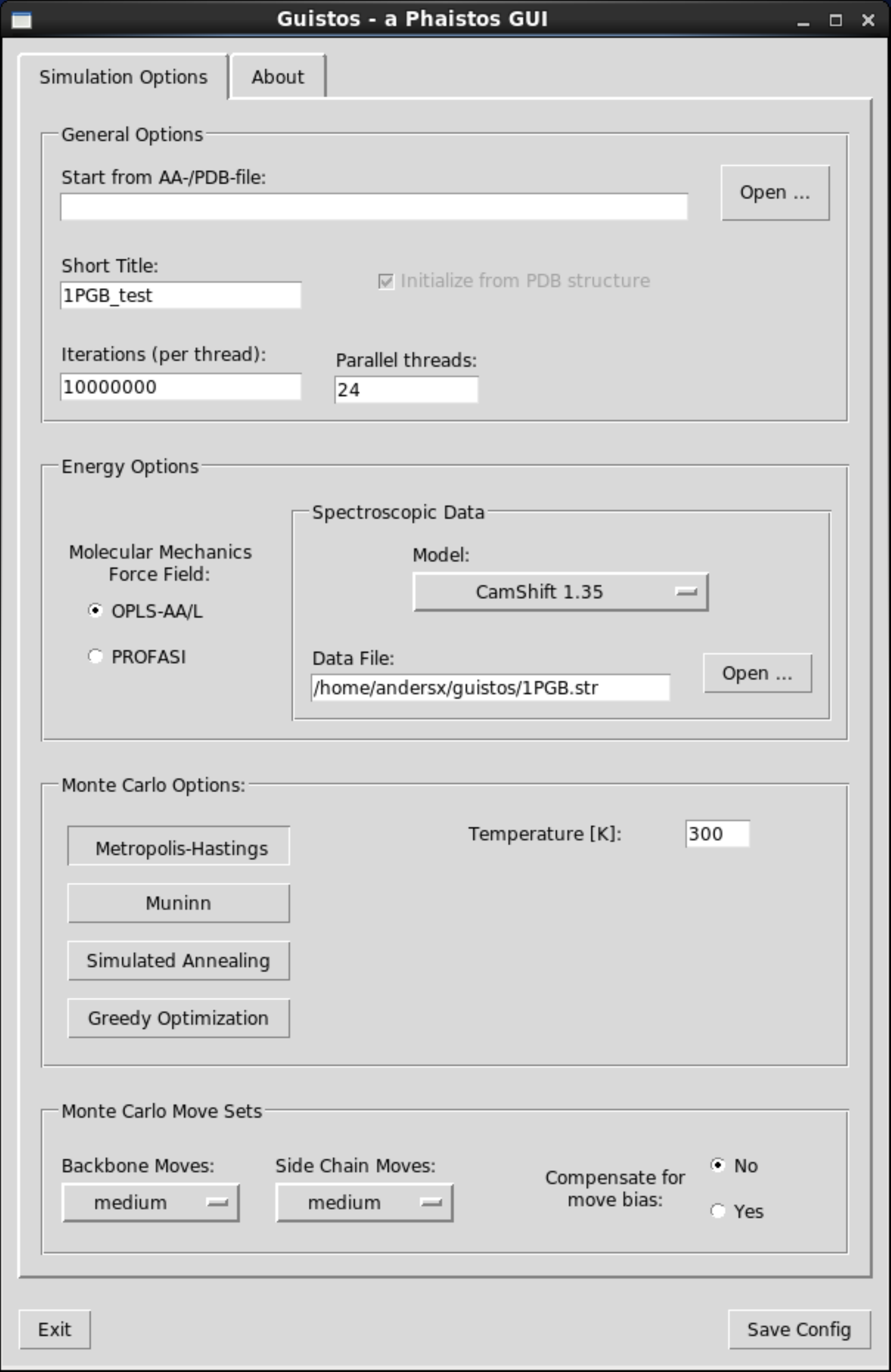}
    \caption{Screenshot of Guistos}
    \label{fig:guistos}
\end{figure}

Using the GUI the user is only presented with the three most basic choices for setting up the simulation.
These are (1) choice of energy terms, (2) type of Monte Carlo simulation and finally (3) a selection of Monte Carlo moves.
A screenshot of Guistos can be seen in Fig.~\ref{fig:guistos}.
Setting up these via Guistos is discussed below.

\subsubsection{Energy Options}

Firstly, the Energy Options section allows the user to select the molecular mechanics force field.
Currently two force fields are supported in PHAISTOS, which are the OPLS-AA/L force field with a GB/SA solvent model, and the PROFASI coarse grained force field.
Use of the PROFASI force field requires the Monte Carlo moves to restraint the bond angle and lengths in the protein to Engh-Huber standard values.
This is automatically done if the PROFASI force field is selected. 
Conversely, the OPLS-AA/L force field includes energy terms for bond angles and lengths and these are degrees of freedom in the simulation if the OPLS-AA/L force field is selected.

Additionally, the Energy Options section allows the user to add restrains from one type spectroscopic data.
Currently energy terms based on CamShift 1.35 and ProCS are supported.
These options requires a NMR-STAR formatted file containing experimental chemical shifts.

\subsubsection{Monte Carlo Options}

This section allows the user to select the four types of Monte Carlo simulation offered by PHAISTOS and the only the most basic options to set up that particular simulation:
Metropolis-Hastings offers the choice of a constant temperature (in Kelvin).
Muninn and Simulated Annealing offer the choice of a temperature range (in Kelvin), and additionally Muninn offers the choice between multicanonical or $1/k$ sampling.
Greedy Optimization does not offer any customizable option. 

\subsubsection{Monte Carlo Move Sets}

Selecting a good mix of the different Monte Carlo moves offered by PHAISTOS can significantly speed up convergence of a simulation, compared to using an inferior move set.
Choosing a good set of moves is in the opinion of this author currently somewhere in between black art and sheer luck, and requires a good deal of experience with simulations in PHAISTOS.

To make it easier for new users, three move sets have been predefined using the experience of this author.
These are named "small", "medium" and "large".
The "small" move set is intended for uses such as refinement or sampling around a compact native state, 
while the "medium" move set is intended for folding simulations that start from extended, but are expected to also sample a native state, 
and finally the "large" move set is intended for sampling conformational space quickly, but will have problems with sampling compact structures.
All move sets sample from TorusDBN (backbone angles) and BASILISK (side chain angles), and an option to remove this bias is also present.

\subsubsection{Using Guistos}

Guistos is freely released under the open source two-clause BSD-license, and can be downloaded from \url{https://github.com/andersx/guistos/}. After specifying all relevant settings in the Guistos window, a configuration-file is saved by pressing the "Save Config" button.
A simulation in PHAISTOS can the be executed via the following command:
\begin{lstlisting}
./phaistos --config-file my_simulation.config
\end{lstlisting}

\chapter{Prediction of Protein Chemical Shifts}

While the relationship between NOE restraints and the underlying protein structure is clear,
the relationship between chemical shifts and the structure is less clear.
Several programs, however, exists which are able to predict protein chemical shifts given a protein structure.
Typically, these chemical shift predictors are parametrized from empirical fits between experimental crystal structures of proteins to their corresponding measured NMR chemical shifts.
Popular programs that employ such empirical include SHIFTX, SPARTA+, SHIFTS and CamShift \cite{SHIFTX,SPARTAPLUS,Osapay1991,CamShift}.
These programs use functional forms that decompose the chemical shift into additive, independent terms.
The accuracy of these fits are inherently limited by the availability and accuracy of empirical data.
A similar program, CheShift, exists, in which the functional forms are interpolated from a large database of QM calculation on representative peptide conformations \cite{CheShift2}.
The authors, however, have not been willing to share the code, but exists as a web-service which allow aplha-carbon and beta-carbon chemical shift calculations.

We have recently explored using quantum mechanics to derive chemical shifts from protein structures. Our amide-proton chemical shift predictor is discussed in our paper \#3 in Appendix A.
Briefly, in the amide proton-only version of ProCS \cite{ProCS}, the chemical shift is calculated as a sum of several independent terms \cite{Parker}:

\begin{equation}
    \delta_\mathrm{H} = \delta_\mathrm{BB}(\phi,\psi) + \Delta\delta_\mathrm{HB} + \Delta\delta_\mathrm{rc}
\end{equation}
where $\delta_\mathrm{BB}(\phi,\psi)$ chemical shift dependence on the backbone angles, $\Delta\delta_\mathrm{HB}$ is a sum over 3 different contributions due to hydrogen bonding and $\Delta\delta_\mathrm{rc}$ is the perturbation due to magnetic field from aromatic side chains \cite{Christensen2011}.
All terms are parametrized by QM methods by fitting the terms to QM calculations on model systems.
The ring current contribution term is discussed in detail in publication \#1 in the appendix.

As we show in the publication, structures generated using amide-proton chemical shift restraints from ProCS have hydrogen bonding geometries that are in substantially better agreement with experimental X-ray structures and back-calculated experimentally measured spin-spin coupling constants, compared to using CamShift as predictor or no chemical shifts in the simulation.
The accuracy of the amide proton-only version of ProCS is lower than SHIFTX, SPARTA+, SHIFTS or CamShift, when experimental protein structures are used as input, but we show that this is likely due to inaccuracies in the experimental coordinates.

Similar to the approximation above, we have made a predictor for all backbone and beta-carbon.
In the backbone atom version of ProCS, the chemical shift is calculated as:
\begin{equation}
    \delta = \delta_\mathrm{BB} + \Delta\delta_\mathrm{HB} + \Delta\delta_\mathrm{rc}
\end{equation}
where $\delta_\mathrm{BB}$ is due to dihedral bond angles in the residue and the neighboring residues, and $\Delta\delta_\mathrm{HB}$ and $\Delta\delta_\mathrm{rc}$ are implemented similarly to those of the amide proton-only version of ProCS.
To accurately calculate the dependence of angles and neighboring residues on carbon and nitrogen chemical shift, we found an accurate description to be:
\begin{equation}
    \delta_\mathrm{BB} = \delta_{i}(\phi_i,\psi_i, \{\chi_i\})
    + \Delta\delta_{i-1}(\phi_{i-1},\psi_{i-1}, \{\chi_{i-1}\})
    + \Delta\delta_{i+1}(\phi_{i+1},\psi_{i+1}, \{\chi_{i+1}\}),
\end{equation}
where $\delta_{i}$ takes into account, the chemical shift due to the $\phi_i,\psi_i$ and $\{\chi_i\}$ angles on the $i$'th residue, and $\Delta\delta_{i-1}$ and $\Delta\delta_{i+1}$ takes into account the perturbation due to the neighboring residue conformation and residue type.

The three terms in $\delta_\mathrm{BB}$ are interpolated through exhaustive scans over all possible conformations of tri-peptides.
To set up the massive number of QM calculations, the FragBuilder Python API was created (see Paper \#4).
FragBuilder is an Python API that makes it possible to easily generate peptide conformations, either via manual definition of dihedral angles or sampling via the BASILISK library \cite{BASILISK}. 
Using the OpenBabel Python API \cite{Babel}, it is furthermore possible to perform molecular mechanics optimizations and write coordinate files in nearly 100 different formats.
FragBuilder provides convenient wrappers and classes for such operations, and only few lines of code are generally needed for generating an input-file.

The FragBuilder Python API was used to generate the more than 2,000,000 peptide structures used to generate the database.
The peptide structures were optimized using the PM6 semi-empirical QM method, and QM chemical shifts were calculated at the OPBE/6-31G(d,p) level using a polarizable continuum model to model an embedding environment.
The resulting tables of chemical shifts were collected and stored in files in Numpy's binary .npz-format \cite{numpy}.

The predictor is programmed into a separate module for PHAISTOS (in C++).
The program loads the Numpy-arrays into the memory and uses existing code to read coordinates and angles.
These tables are roughly 10GB for each nucleus, so the current version of ProCS requires about 64GB of RAM for predicting backbone atom and beta-carbon chemical shifts efficiently.

\section{Initial Results}
The code is currently not ready for use in simulations, other than for testing purposes, due to the massive memory requirements, and parallelization is not yet complete, so no results in this respect can be presented here.

Initial test show that calculating chemical shifts via the ProCS module is about 5 times faster than the CamShift energy term in PHAISTOS and roughly same speed as the PROFASI energy term.
Note, that the CamShift and PROFASI energy terms use a caching algorithm which effectively means that only terms that depend on atoms that are move during a Monte Carlo move have to be re-calculated each move.
An initial cached version of ProCS is around 5 times faster than the non-cached version, and thus faster than the coarse-grained PROFASI force field. Fast evaluation of chemical shifts is crucial for including the chemical shift predictor in the energy function when simulating folding of larger proteins ($>100$ amino acids), where the CamShift predictor is currently too slow for our purpose.

We have assessed the accuracy of ProCS for alpha-carbon and beta-carbon atoms by comparison to benchmark QM calculations on an entire proteins.
The experimental structures of Protein G and Ubiquitin (PDB-codes: 2OED and 1UBQ, respectively) were protonated using the PDB2PQR webinterface \cite{pdb2pqr1,pdb2pqr2}.
Additional structures were generated by minimizing the X-ray structures in Tinker with the AMBER, CHARMM22/CMAP and AMOEBA force fields with a GB/SA solvent model.
The chemical shifts of the resulting structures were calculated in GAUSSIAN 09 \cite{g09} at the OPBE/6-31G(d,p) level with a polarizable continuum solvent model. The results are summarized in table \ref{tab:procs_results}.

The QM calculations on Ubiquitin are in slightly better agreement with the ProCS predicted number, than the CheShift and CamShift predicted values, based on RMSD and $r^2$ values. For Protein G CheShift are and CamShift RMSD values are slightly lower for alpha-carbon, while ProCS has a lower RMSD for beta-carbon.
The general trend is that the predictors are comparable in accuracy.

\begin{table}[h]
    \caption{Comparison of agreement between QM calculation of alpha-carbon and beta-carbon chemical shifts and predicted chemical shifts, for X-ray structures of Ubiquitin and Protein G, and structures minimized with the AMBER, CHARMM22/CMAP and AMOEBA force fields.}
    \begin{center}
    \begin{threeparttable}
    \begin{tabular}{l r r r r r r}
               & ProCS  &       &  CheShift&      &  CamShift& \\
CA/Ubiquitin   & $r^2$  & RMSD  &  $r^2$   &RMSD  &  $r^2$   &RMSD\\\hline
1UBQ (X-ray)   & 0.754  & 2.54  &  0.697   &3.63  &  0.666   &2.97\\
AMBER          & 0.815  & 1.93  &  0.789   &3.19  &  0.763   &2.41\\
CHARMM22/CMAP  & 0.897  & 2.78  &  0.775   &2.12  &  0.827   &2.68\\
AMOEBA         & N/A    & N/A   &  0.851   &3.94  &  0.886   &2.26\\
 &&&&&&\\
CA/Protein G   & $r^2$  & RMSD  &  $r^2$   &RMSD  &  $r^2$   &RMSD\\\hline
2OED (X-ray)   & 0.894  & 2.37  &  0.883   &1.66  &  0.887   &2.21\\
AMBER          & 0.824  & 3.02  &  0.883   &1.87  &  0.883   &1.87\\
CHARMM22/CMAP  & 0.907  & 2.60  &  0.814   &2.13  &  0.839   &2.82\\
AMOEBA         & 0.914  & 1.90  &  0.866   &3.84  &  0.755   &2.82\\
 &&&&&&\\
CB/Ubiquitin   & $r^2$  & RMSD  &  $r^2$   &RMSD  &  $r^2$   &RMSD\\\hline
1UBQ (X-ray)   & 0.947  & 3.44  &  0.945   &3.90  &  0.941   &3.58\\
AMBER          & 0.983  & 1.91  &  0.965   &2.85  &  0.964   &2.54\\
CHARMM22/CMAP  & 0.980  & 2.76  &  0.971   &5.22  &  0.970   &3.34\\
AMOEBA         & N/A    & N/A   &  0.957   &6.34  &  0.950   &4.30\\
 &&&&&&\\
CB/Protein G   & $r^2$  & RMSD  &  $r^2$   &RMSD  &  $r^2$   &RMSD\\\hline
2OED (X-ray)   & 0.992  & 2.87  &  0.983   &2.2   &  0.983   &3.10\\
AMBER          & 0.974  & 2.91  &  0.982   &2.63  &  0.982   &2.63\\
CHARMM22/CMAP  & 0.991  & 2.68  &  0.979   &4.95  &  0.985   &3.08\\
AMOEBA         & 0.984  & 3.83  &  0.977   &6.29  &  0.977   &4.06\\
    \end{tabular}
    \end{threeparttable}
    \end{center}
    \label{tab:procs_results}
\end{table}


\chapter{Determined protein structures}
\label{chapter:results}
This section describes all test-targets which I have attempted to fold using the methodologies presented in the previous chapters.
All protein structures, chemical shift and NOE data used in this thesis is available from \url{https://github.com/andersx/cs-proteins/}.

\section{Barley Chymotrypsin Inhibitor II}

An especially interesting target in this study is the barley chymotrypsin inhibitor II (CI-2). CI-2 is a 63 residue protein which consists of an $\alpha$-helix which connects via a very flexible handle to a small $\beta$-sheet region.

The chemical shifts data supplied by Kaare Theilum (personal communication) was obtained using a fully automated procedure.
The ADAPT-NMR \cite{adaptnmr} protocol was used to record all necessary NMR data and automatically assign the chemical shifts.
Data collection and assignment was completed in only 11 hours with minimal human intervention.
As we demonstrate, a structure could be determined computationally from these chemical shifts in only two days running on 12 cores.

\subsection{Computational methodology}
Several folding protocols were tried for this protein. All runs were performed as 72 independent trajectories which ran for 50 mio MC steps (iterations).
Sampling was carried out using either TorusDBN or TorusDBN-CS to bias the backbone moves and the PROFASI force field was used in all simulations.
One simulations used an experimental version of TorusDBN-CS, supplied by Lars Bratholm, which was trained on only high-resolution X-ray structures (available from \url{https://github.com/andersx/cs-proteins/}).
Three simulations used an energy function based on CamShift using a cauchy distribution with variable $\gamma$-weight as energy function.
Additionally, three simulations used a potential on the radius of gyration to restrict the sampling to only compact structures \cite{Phaistos_oldest}.
Sampling was performed in the multicanonical ensemble with a thermodynamic beta-range from 0.6 to 1.1, corresponding to a temperature range of 272K to 500K. The MC move set was comprised of 40\% CRISP moves, 10\% pivot moves and 50\% uniform side chain moves.

\begin{table}[h]
    \caption{Protocols used in the folding of the CI-2 protein and success rates.}
    \begin{center}
    \begin{threeparttable}
    \begin{tabular}{l l l l l l}
        Sampling        & Force Field   & CS Energy         & Correct fold\tnote{a} & Iterations/day\tnote{b}\\\hline
          TORUS-CS + RG\tnote{c} & PROFASI       & CamShift          & 13            & $10 \times 10^6$ \\
          TORUS-CS      & PROFASI       & CamShift          & 15            & $11 \times 10^6$\\
          TORUS         & PROFASI       & CamShift          & 0             & $11 \times 10^6$\\
 TORUS-CS + PP\tnote{c} & PROFASI       & None              & 4\tnote{d}      & $49 \times 10^6$\\
 TORUS-CS\tnote{e}+ RG\tnote{c} & PROFASI      & None              & 0             & $49 \times 10^6$\\
          TORUS-CS      & PROFASI       & None              & 0             & $49 \times 10^6$\\
          TORUS         & PROFASI       & None              & 0             & $49 \times 10^6$\\
    \end{tabular}
    \begin{tablenotes}
        \item[a] Number of threads with a CA-RMSD of $<5$ \AA\ (using all residues).\\
        \item[b] Numbers are \textit{per} thread.\\
        \item[c] RG denote the use of an additional radius of gyration potential.\\
        \item[d] Structures with the lowest energy did not correspond to the native structure in this run.\\
        \item[e] This run was carried out using TorusDBN-CS trained using only high-quality X-ray structures.
    \end{tablenotes}
    \end{threeparttable}
    \end{center}
    \label{tab:ci2}
\end{table}
\subsection{Folding results}
\label{sec:ci2_results}
Three of the 7 attempted simulation types sample structures close to the experimental X-ray structure 1YPA (here loosely defined as a CA-RMSD $<5$ \AA\ for all CA atoms. Results are summarized in table \ref{tab:ci2}.
Only simulations using chemical shift biased sampling through TorusDBN-CS are able to sample the correct fold.

Furthermore, it was noted, that simulations that sample from either TorusDBN or TorusDBN-CS with only the PROFASI force field as energy function do not generate compact structures.
To overcome this deficiency, additional simulations were carried out using a radius of gyration potential.
In the case of sampling from TorusDBN-CS, the radius of gyration potential is enough to get a few samples with the correct fold. Here four of 72 threads would generate the correct fold, but unfortunately the lowest energy structures were found around 8-11 \AA\ CA-RMSD. Evidently, the PROFASI force field alone is not accurate enough to describe the native CI-2 structure.
Three simulations were performed with an energy term based on CamShift in addition the PROFASI force field.
Demonstrably, the increased accuracy from a better energy function cause increased sampling around the native state.

Due to a very flexible region of CI-2 (residues 33 to 42), and somewhat flexible tails the residue range used to calculate CA-RMSD values is restricted to residue 4-34,43-63 in the following.
All runs were carried out on 3 24-core AMD Opteron 6172 servers running at 2.1 GHz. 

A run similar to the most successful was also run carried out on a faster a 12-core Intel X5675 node running at 3.07 GHz (using new random seeds).
PHAISTOS input to reproduce these folding simulation is given below.

\begin{lstlisting}
./phaistos --aa-file ci2.aa \
  --iterations 50000000 \
  --threads 12 \
  --monte-carlo-muninn 1 \
  --monte-carlo-muninn-min-beta 0.6 \
  --monte-carlo-muninn-max-beta 1.1 \
  --monte-carlo-muninn-independent-threads 1 \
  --monte-carlo-muninn-weight-scheme multicanonical \
  --backbone-dbn-torus-cs 1 \
  --backbone-dbn-torus-cs-initial-nmr-star-filename ci2.str \
  --energy-profasi-cached 1 \
  --energy-camshift-cached 1 \
  --energy-camshift-cached-star-filename ci2.str \
  --energy-camshift-cached-energy-type 11 \
  --move-backbone-dbn 1 \
  --move-backbone-dbn-weight 0.1 \
  --move-backbone-dbn-implicit-energy 1 \
  --move-crisp-dbn-eh 1 \
  --move-crisp-dbn-eh-weight 0.4 \
  --move-sidechain-uniform 1 \
  --move-sidechain-uniform-weight 0.5
\end{lstlisting}

\begin{figure}%
    \centering
    \includegraphics[width=0.5\textwidth]{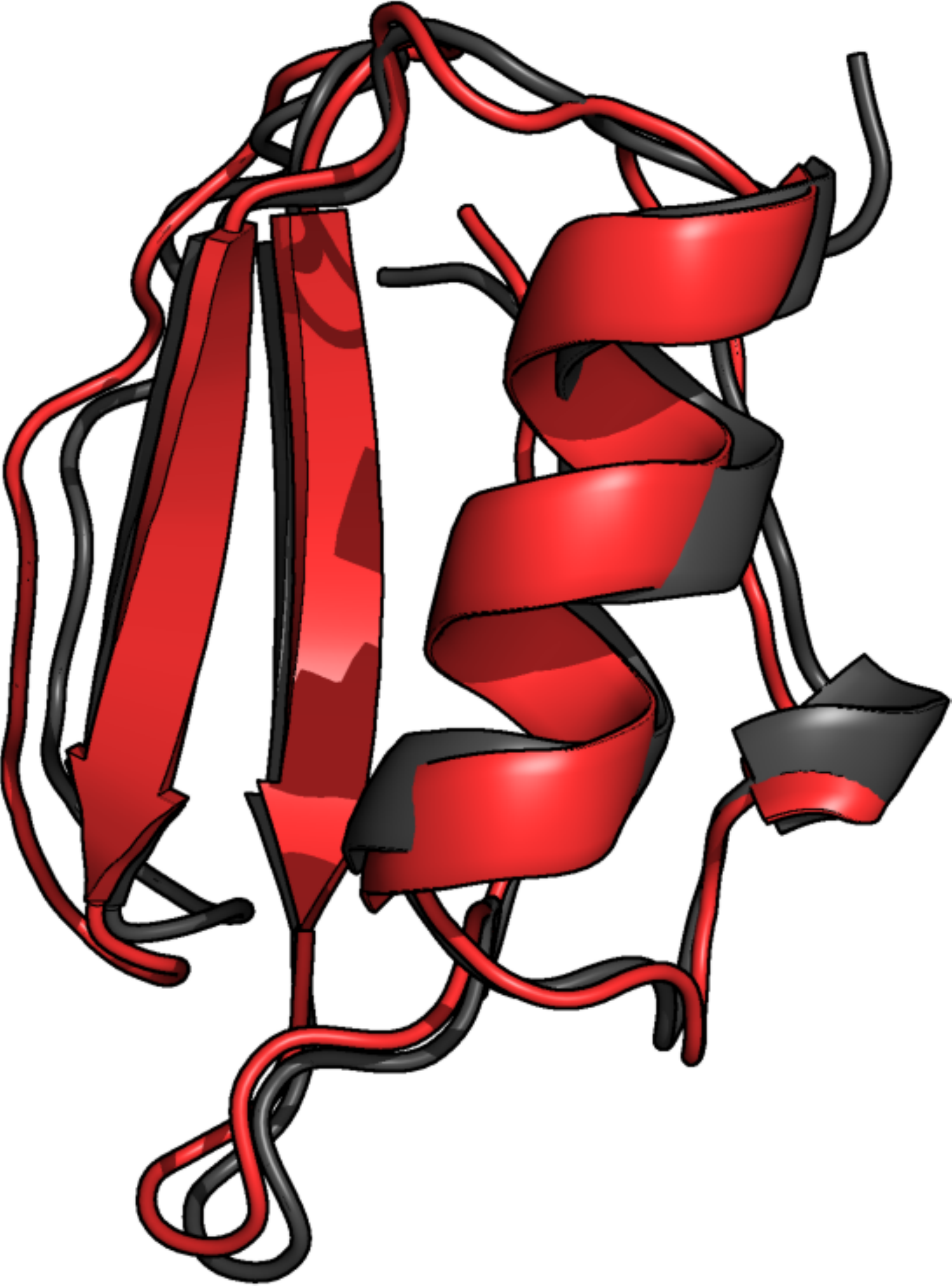}
    \caption{The CI-2 structure refined to 1.1 \AA~by Lars Bratholm. The refinement was carried out by including a multibody-multinomial poteintial of mean force in the simulation.}
    \label{fig:ci2_lars}%
\end{figure}

This simulation took two days, with a total of 2 out of 12 threads successfully identifying the native structure as having the lowest energy.
This simulation yielded a lowest energy structure a 2.76 \AA~CA-RMSD from the X-ray structure, and a lowest RMSD structure at 1.11 \AA. Later, this lowest energy sample was further refined by Lars Bratholm to a CA-RMSD of only 1.1 \AA~using an additional multibody-multinomial potential of mean force in the energy function \cite{mumu}. This refinement simulation took 24 hours on 8 cores.
This structure is displayed in Fig.~\ref{fig:ci2_lars}.

In conclusion, the data for CI-2 was recorded in merely 11 hours via a fully automated process. A structure comparable to conventional NMR structures could then be determined after 36 hours. After an additional 24 hours, a structure that rivals X-ray structures was further determined by Lars Bratholm.

\begin{figure}%
    \centering
    \subfloat[NMR structure (red)]{
        {\includegraphics[width=0.3\textwidth]{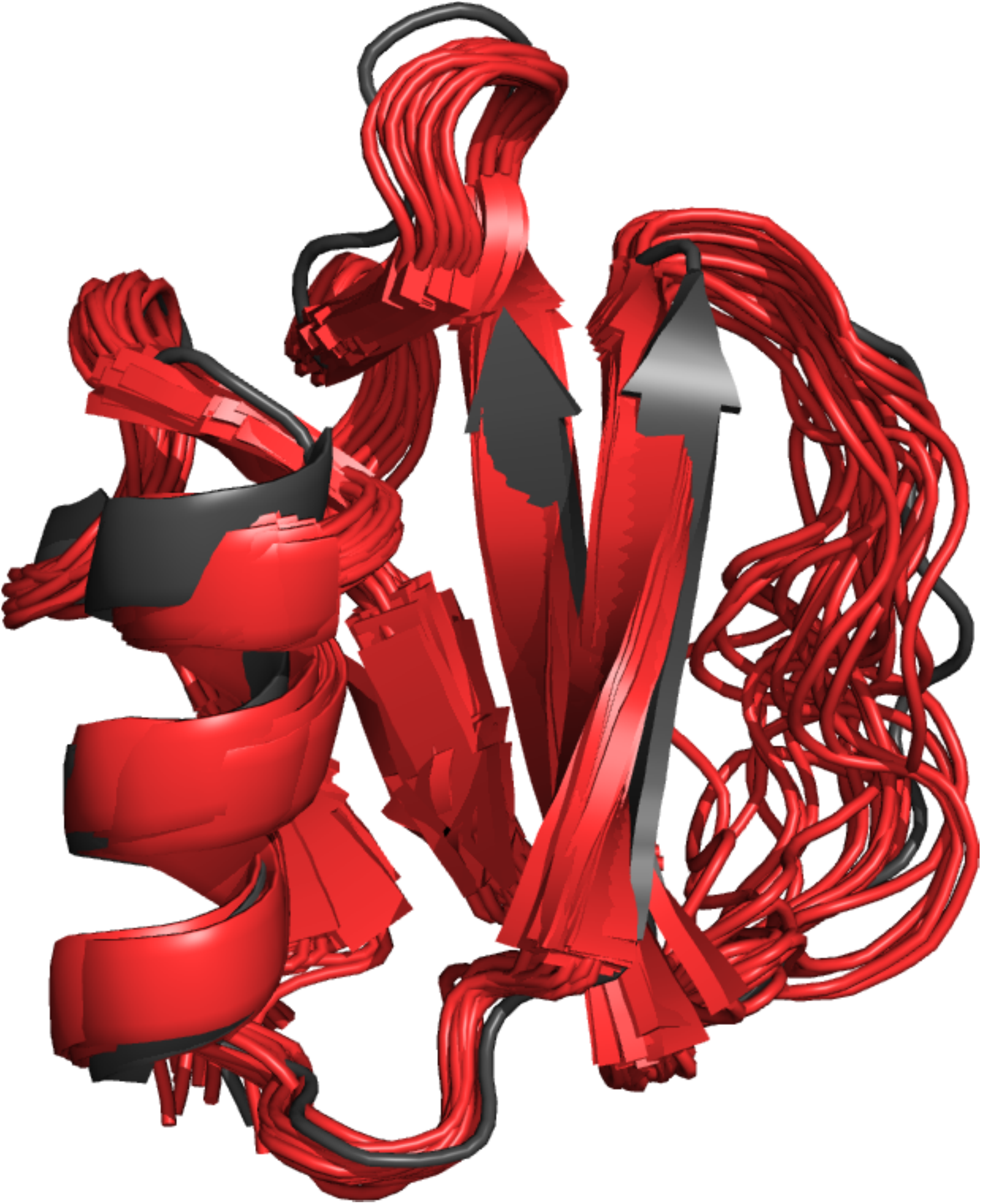}}
    }
    \subfloat[Lowest RMSD structure (blue)]{
        {\includegraphics[width=0.3\textwidth]{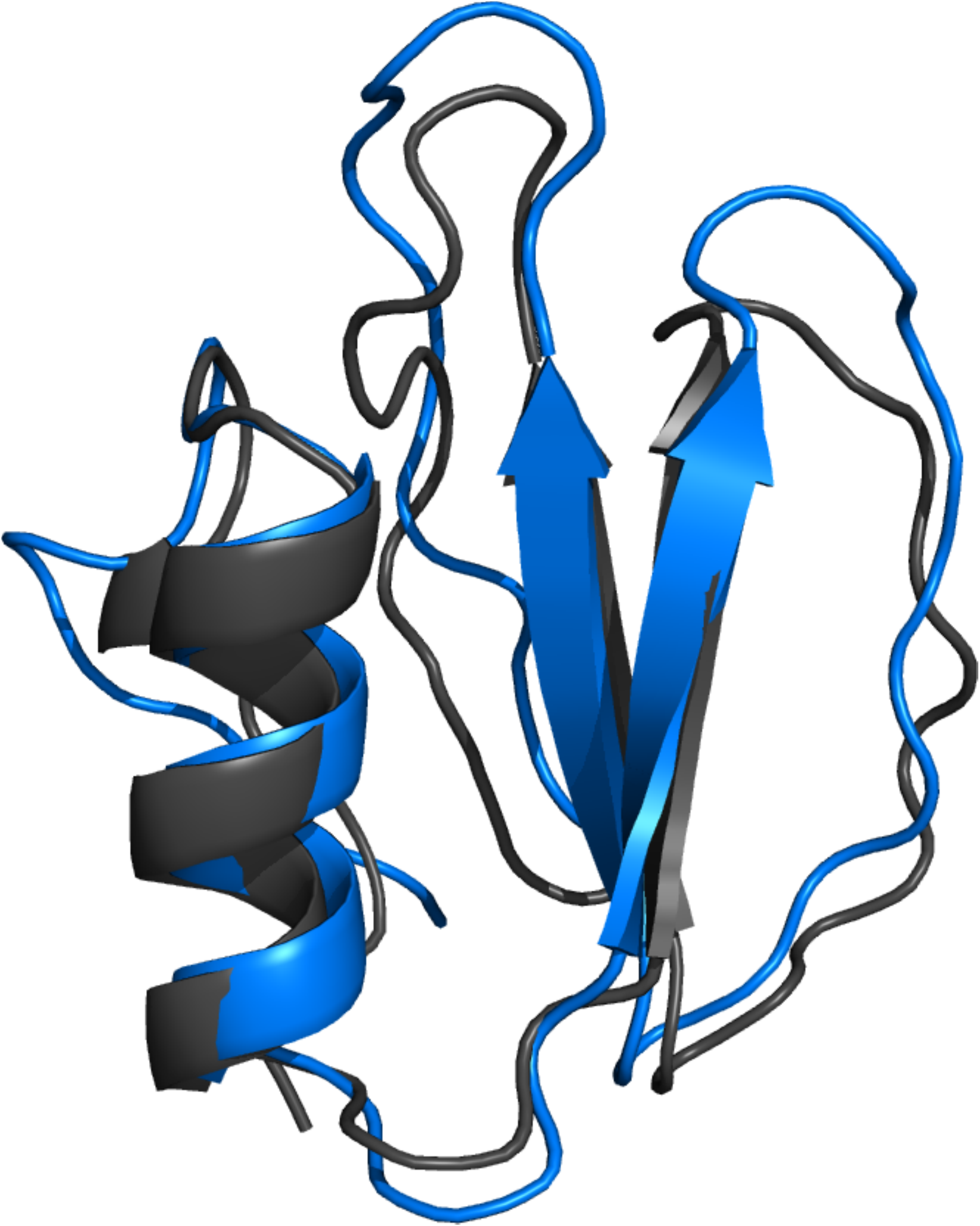}}
    }
    \subfloat[Lowest energy structure (green)]{
        {\includegraphics[width=0.3\textwidth]{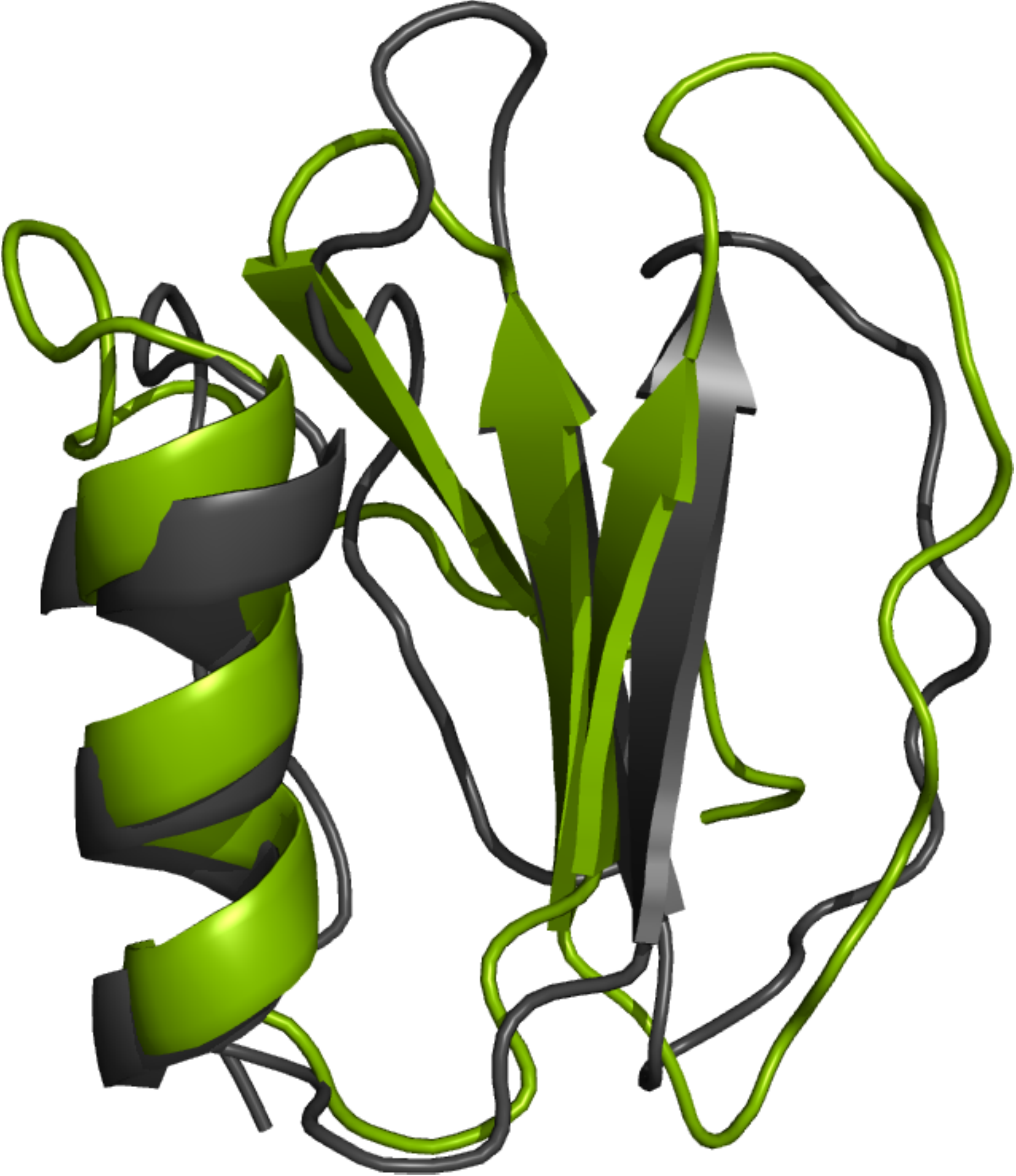}}
    }
    \caption{Structures compared to the X-ray structure 1YPA. All structures are aligned using the residues 12-32,43-52. (a) shows the 3CI2 structure NNR structure. Note the flexible domain which is excluded from the fit-range. (b) Shows the lowest RMSD structure (1.113 \AA RMSD). (c) shows the lowest energy sample (2.76 \AA RMSD). }
    \label{fig:ci2}%
\end{figure}

\clearpage
\section{Folding of small proteins (\textless 100 AA)}

\begin{figure}
    \centering
    \subfloat[Protein G, 1.0 \AA.]{
        {\includegraphics[width=0.45\textwidth]{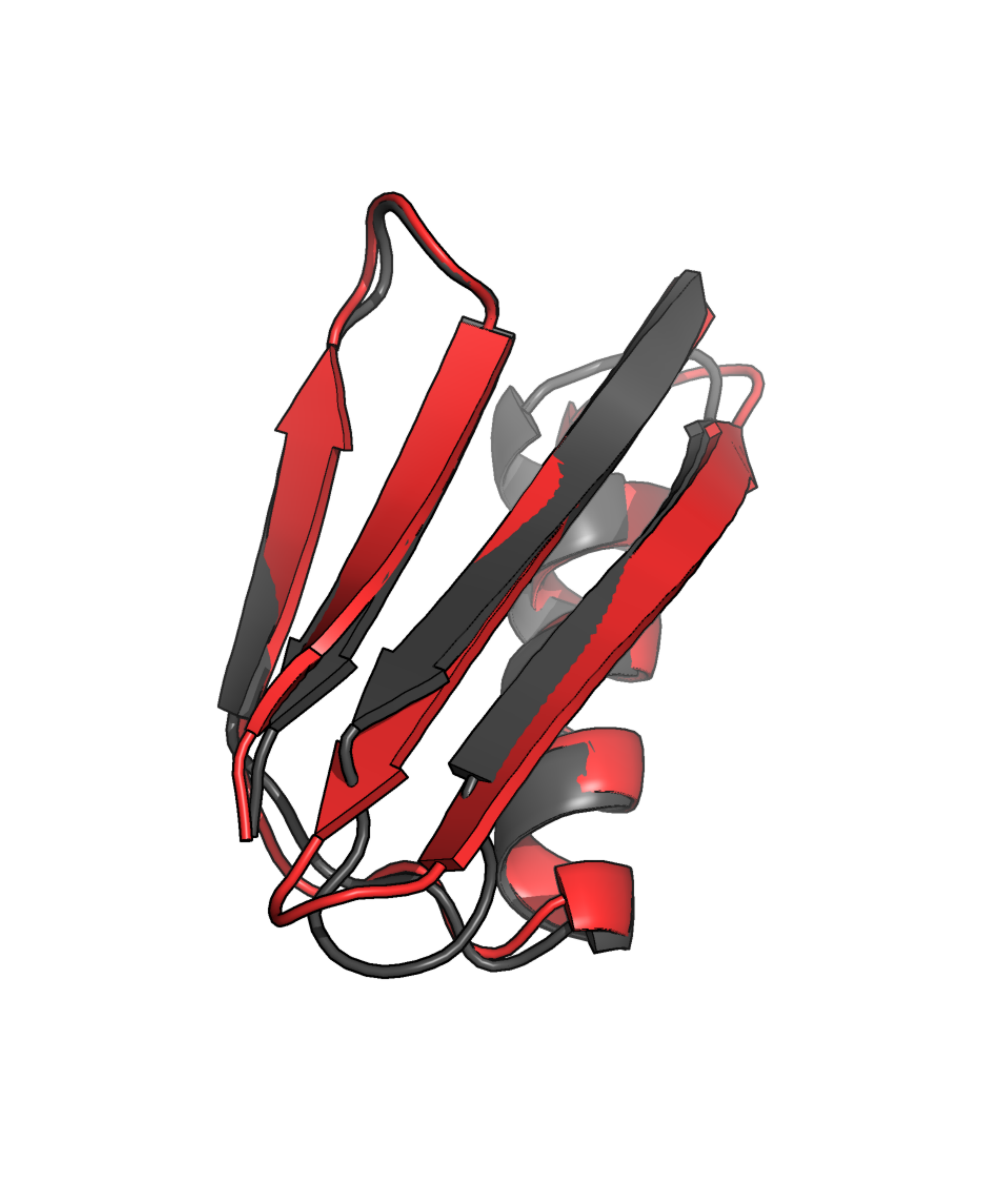}}
    }\quad
    \subfloat[Engrailed Homeodomain, 1.1 \AA.]{
        {\includegraphics[width=0.45\textwidth]{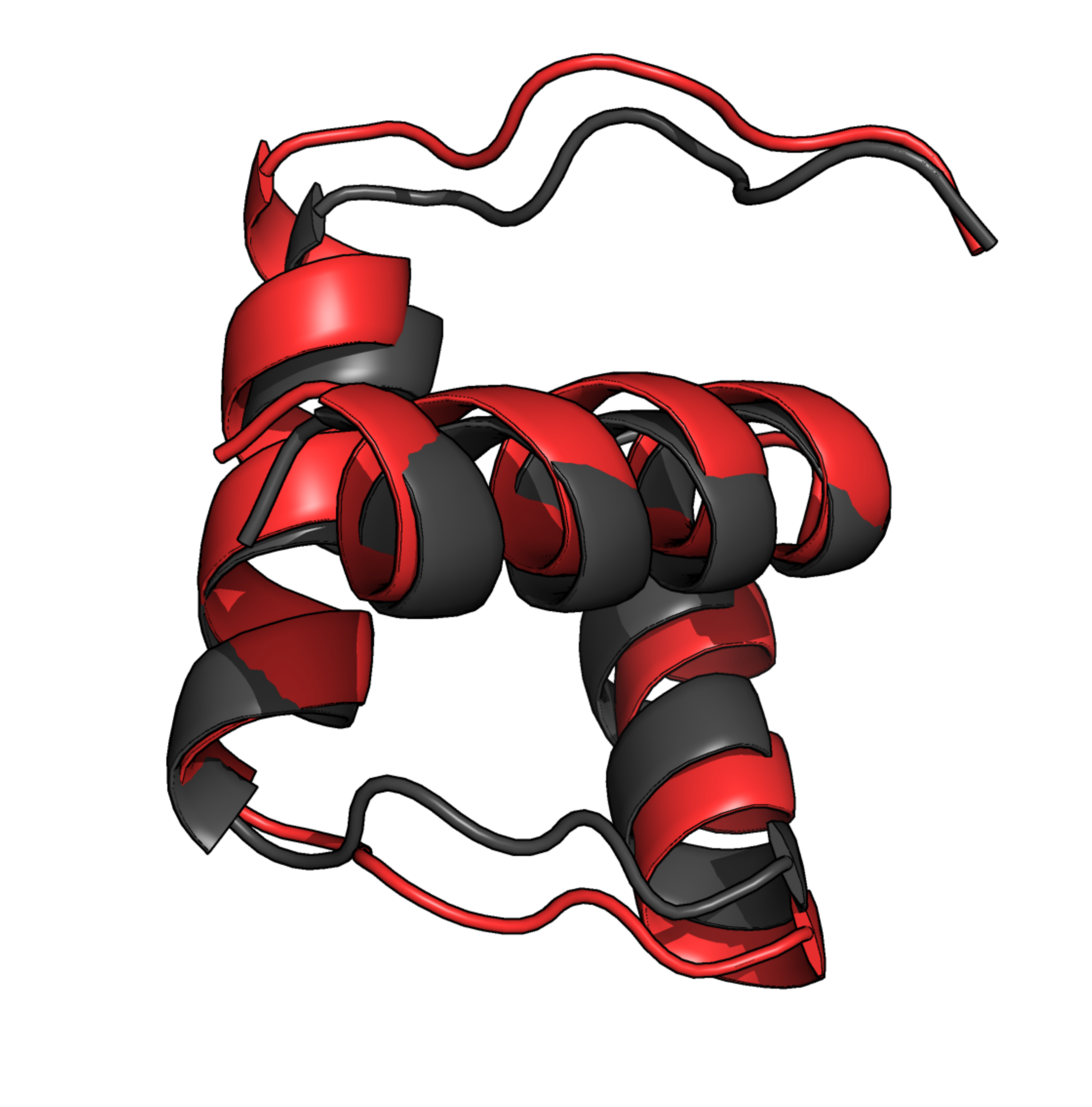}}
    }\\
    \subfloat[FF Domain, misfold.]{
        {\includegraphics[width=0.45\textwidth]{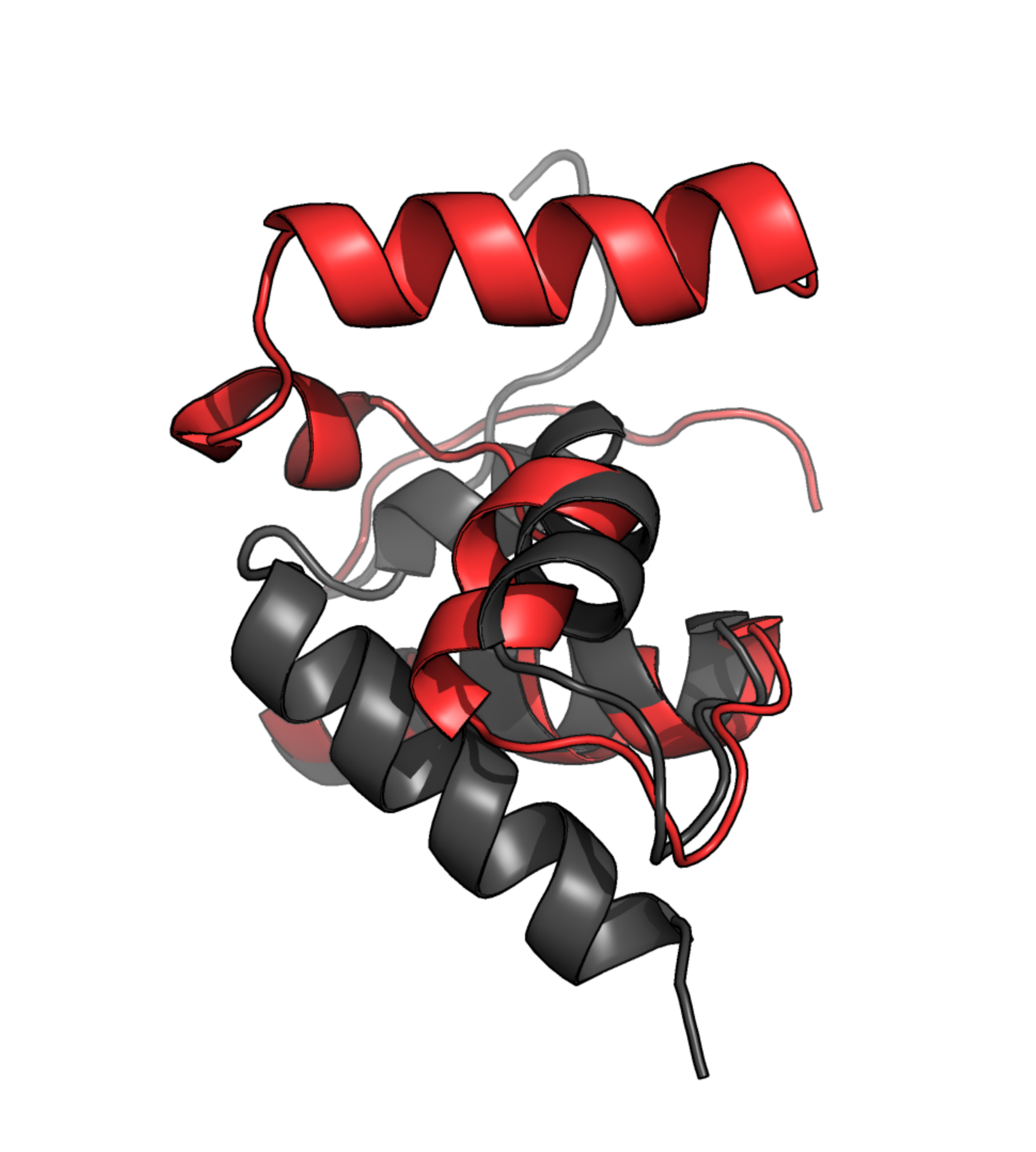}}
    }\quad
    \subfloat[Ubiquitin, 3.8 \AA]{
        {\includegraphics[width=0.45\textwidth]{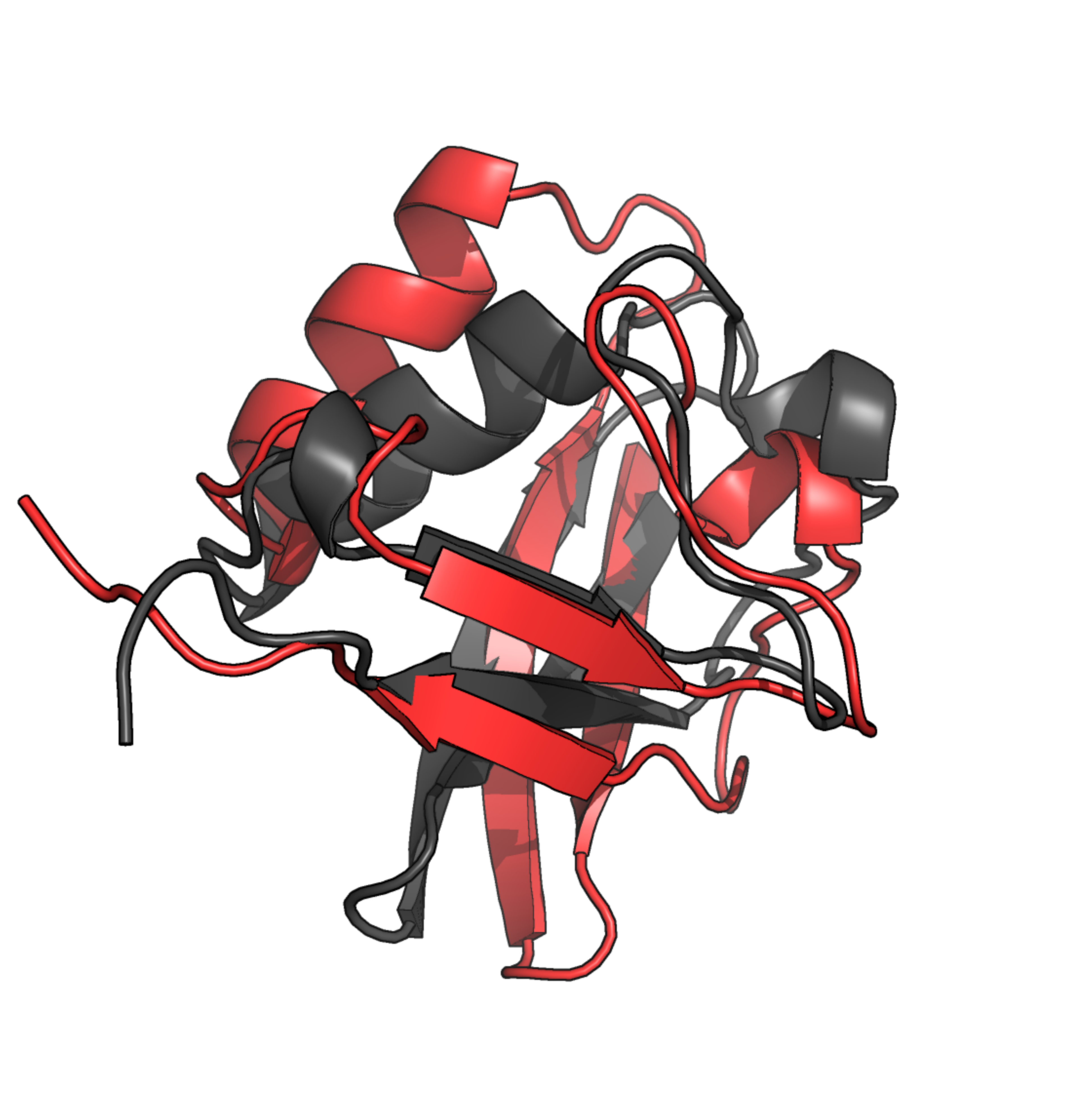}}
    }\\
    \caption{The lowest energy structures found for four different proteins (red). Superimposed on corresponding X-ray structures (grey). The FF Domain structures in (c) is aligned using only residues 1-40 to emphasize the misfold.}
    \label{fig:small_pics}%
\end{figure}

A test set of 5 small proteins were folded using the code. The results are summarized in table \ref{tab:folding_small}. The test set is a diverse set of structures with different contents of alpha-helix and beta-sheet conformations. 
The settings are similar to the ones used to fold the CI-2 structure mentioned in the previous section, except that the Protein G, Ubiquitin, FF Domain and Engrailed Homeodomain (ENHD) simulations used a chemical shift energy based on a Gaussian distribution with fixed weights (--energy-camshift-cached-energy-type 3), and not based on a Cauchy distribution (--energy-camshift-cached-energy-type 11).
\begin{table}
    \caption{The five small proteins folded using the setup presented in this section, and their RMSD for the lowest energy sample.}
    \begin{center}
    \begin{threeparttable}
    \begin{tabular}{l l l l l  l l}
Name                & Lengh    & Type & PDB     & RefDB     & RMSD-range    & Final RMSD   \\\hline
Protein G           & 56       & a/b & 2OED    & 2575      & All           & 1.0           \\
Engrailed Homeodomain & 61     & B   & 1ENH    & 15536     & 8-53          & 1.1           \\
FF Domain           & 71       & a/b & 1UZC    & 5537      & 11-67         & 10.2         \\
Ubiquitin           & 76       & a/b & 1UBI    & 17769     & 1-70          & 3.8           \\
CI-2                & 63       & a/b & 1YPA    & N/A\tnote{a}& 4-34,43-63  & 2.6\tnote{b}
    \end{tabular}
    \begin{tablenotes}
    \item[a] Using automatically assigned data obtained from Kaare Theilum (personal communication -- see \url{https://github.com/andersx/cs-proteins/}).
    \item[b] The number reported is discussed in section \ref{sec:ci2_results}.
    \end{tablenotes}
    \end{threeparttable}
    \end{center}
    \label{tab:folding_small}
\end{table}
Total energy was calculated as the PROFASI force field energy plus the CamShift energy term based on a Gaussian distribution with fixed weights plus the likelihood from TorusDBN-CS.
Protein G and ENHD structures could be determined very reliably to CA-RMSDs of 1.0 \AA~and 1.1 \AA~from the experimental structures, respectively. The lowest energy structures are presented in Fig.~\ref{fig:small_pics}a and \ref{fig:small_pics}b.

For the FF Domain, a folded state with a lower energy than the native state was located.
A state corresponding to the correct fold was consistently being sampled in most threads, but the lowest energy stat was a misfold, where an alpha-helix towards the C'-end is packed wrongly.
This result suggests, that the combination of the PROFASI force field and the chemical shift energy from CamShift and TorusDBN-CS does not always discriminate the potential energy surface with sufficient accuray. The energy from CamShift (and thus the chemical shift RMSD values, since the energy function was a Gaussian distribution) was comparable between samples around the correct fold and the lowest energy mis fold. The lowest RMSD structre (3.2 \AA) had a CamShift energy of 803 kcal/mol, while the lowest energy structure had a CamShift energy of 797 kcal/mol.
The lowest energy misfold is displayed in Fig.~\ref{fig:small_pics}c.
In the Ubiquitin simulations, the lowest energy conformations were not in exceptional agreement with the experimental structure with a CA-RMSD of 3.8 \AA - see Fig.~\ref{fig:small_pics}d.
 Again, this must be attributed to lack of "funneling" of the energy landscape around the native state, since sampling evidently is performed close to this state.

Collectively, these result show, that sampling from TorusDBN-CS in PHAISTOS is indeed very efficient, but better energy functions are required in some cases. In one case, the CamShift energy term had a lower energy by 6 kcal/mol for a misfold, than for a sample close to the native state.
Another option would be using a better molecular mechanics force-field. PHAISTOS already supports the OPLS-AA/L but using this would increase simulation times by more than one order of magnitude, and would be unacceptable for simulations on larger structures.

\clearpage

\section{Folding of larger proteins (\textgreater 100 AA)}

This section presents folding results on a set of larger proteins (\textgreater 100 AA) with known structures.
It is worth to note, that using sparse NMR data, only three structures \textgreater 200 residues have been determined: Alg13 (201 AA), Rhodopsin (225 AA) and MBP (376 AA) using the ROSETTA program with the "resolution-adapted structural recombination" (RASREC) protocol \cite{Lange2012,LangePNAS2012}.

Alg13 was solved using backbone chemical shifts, and only 52 NOE restraints, to an CA-RMSD of 4 \AA~to the experimental NMR structure.
Rhodopsin was folded to an CA-RMSD of 1.9 \AA~to the X-ray structure using 215 NOE restraints, backbone chemical shifts chemical shifts and RDCs.
The MBP protein is a two-domain protein of 376 residues.
MBP was folded to an RMSD of 3.6 \AA~using 1235 NOE restraints, backbone chemical shifts chemical shifts and RDCs.
The NOEs corresponded to 55\% yield of restraints, which, for the most part,  were not automatically assigned.
An attempt to use only automatically assigned NOEs yielded 455 restraints, which corresponds to a yield of 20\%. Using these, however, the MBP structure could only be determined to a total CA-RMSD of 12.3 \AA.
The N-terminal domain was converged to 2.7 \AA, but the C-terminal domain and the angle between the two domains was incorrectly folded.

Langer \textit{et al.} have demonstrated that by using a special side-chain labeling scheme a few NOE restraints (around 150-250) can be automatically assigned, and these are generally enough to fold the structures using a ROSETTA protocol\cite{LangePNAS2012}.
The scheme is a "ILV-labeling" scheme, where the methyl groups of isoleucine, leucine and valine side-chains are selectively labeled with $^{13}$C and $^1$H isotopes. These groups are commonly found in the core region of the protein and these methyl groups will generally be in contact with each other, thus being able to provide valuable NOE distance restraints. This corresponds to only assigning 10-20\% of the full spectrum.

From the structures in the study by Langer \textit{et al.}, only five structures consist of one chain only, and only those could be simulated in PHAISTOS.
These five structures were selected into the test-set used here, and additionally Prolactin and the Top7 proteins were added.
The ILV-data used by Langer \textit{et al.} could only be obtained through correspondence with the authors for Rhodopsin. 
For all other proteins, synthetic NOE contacts were generate by simulating a synthetic spectrum.
An overview of the proteins and the number of synthetic NOE restraints can be found in Table \ref{tab:folding_large}.

\begin{table}[h]
    \caption{Folded structure.}
    \begin{center}
    \begin{threeparttable}
    \begin{tabular}{l l l l r l r l}
Name                & Lengh    & Type   & PDB     & BMRB    & RMSD-range    & \#NOEs &  RMSD  [\AA]\\\hline
Top7                & 120      & a/b    & 2MBL    & 19404     & 5-104       &  62   & 2.1 \\
MSRB                & 151      & a/b    & 3E0O    & 17008     & 36-105      & 170   & N/A \\
WR73                & 183      & a/b    & 2LOY    & 16833     & 1-36,66-181 & 215   & N/A \\
HR4660B             & 174      & a/b    & 2LMD    & 1870      & 16-162      & 68    & N/A \\
Rhodopsin           & 219      & B      & 2KSY    & 16678     & All         & 195   & 2.5 \\
Prolactin           & 199      & B      & 1RWS    & 5599      & 6-183       & 68    & 3.5 \\
Savinase            & 269      & a/b    & 1WVN    & Note\tnote{a,b} & Note\tnote{b} & 270 & 2.9 \\
MBP                 & 376      & a/b    & 1EZ9    & 6807      & All         &  1054 & N/A
    \end{tabular}
    \begin{tablenotes}
    \item[a] Evolutionary distance constraints from the EVFold were used in this case.
    \item[b] Available from: \url{http://github.com/andersx/cs-proteins/}
    \end{tablenotes}
    \end{threeparttable}
    \end{center}
    \label{tab:folding_large}
\end{table}

\subsection{Folding protocol}

The folding simulation settings were similar to those used to fold small proteins, with the exception that the CamShift energy term was too slow to be used in practice.
The additional NOE distance restraint term used a flat-bottom potential with a width of 4 \AA~around the equilibrium distance, and a quadratic potential outside this range.
This was done using the existing NMR inference module in PHAISTOS.

However, using this potential turned out to be quite problematic. 
Once a distance restraint was fulfilled, the simulation would in most cases never break the contact again.
Consequently, an empirical factor of 1/128 was multiplied onto the NOE energy.
This factor was determined by running simulations on the Top7 structure with weights from $1/2^1$ to $1/2^{10}$
Unfortunatly, due to this problem, no good structures for MSRB, WR73, HR4660B and MBP could be located. 
After a few 1,000,000 steps the structures located local minima which fulfilled a number of distance restraints, but it was impossible to escape these minima.
The Top7 structure folded to an RMSD of 2.1 \AA.
This result, however, is not surprising, since Top7 has been shown to fold using only the PROFASI force field.
The Prolactin and Rhodopsin structures converged to structures at 8.5 and 7.8 \AA~RMSD from the X-ray structures.
\\\\The settings to run the simulations are displayed below:
\begin{lstlisting}
./phaistos --aa-file rhodopsin.aa \
  --iterations 50000000 \
  --threads 72 \
  --monte-carlo-muninn 1 \
  --monte-carlo-muninn-min-beta 0.6 \
  --monte-carlo-muninn-max-beta 1.1 \
  --monte-carlo-muninn-independent-threads 1 \
  --monte-carlo-muninn-weight-scheme multicanonical \
  --backbone-dbn-torus-cs 1 \
  --backbone-dbn-torus-cs-initial-nmr-star-filename \
                                      rhodopsin.str \
  --energy-profasi-cached 1 \
  --energy-isd-dist 1 \
  --energy-isd-dist-likelihood square_well \
  --energy-isd-dist-data-filename noe_ilv.txt \
  --energy-isd-dist-sample-gamme 0 \
  --energy-isd-dist-sample-sigma 0 \
  --energy-isd-dist-weight 0.0078125 \
  --move-backbone-dbn 1 \
  --move-backbone-dbn-weight 0.08 \
  --move-backbone-dbn-implicit-energy 1 \
  --move-crisp-dbn-eh 1 \
  --move-crisp-dbn-eh-weight 0.42 \
  --move-sidechain-uniform 1 \
  --move-sidechain-uniform-weight 0.5
\end{lstlisting}

\subsection{Refinement protocol}

Due to the low efficiency of the NOE code for large structures, a new NOE module was written for PHAISTOS. In this module, the potential from the ROSETTA RASREC protocol was used \cite{LangePNAS2012}.
In brief, this is also a flat-bottom potential, but with a linear penalty, rather than quadratic, outside the flat area.
This was done in order to allow more contacts to be broken throughout the simulation in order to enhance conformational sampling.
Additionally, the module only has a certain fraction of all restraints active at a time.
A Monte Carlo move was created which turned off one random, active NOE restraint and activated one random, deactivated restraint.
The resulting energy difference was subtracted as a move-bias, in order to force a 100\% acceptance rate for this move.
This was done, because the energy difference between and active restraint (which is usually close to zero) and an inactive restraint (usually a large number) caused this move to have a low acceptance rate.

Using the new NOE module, a refinement on the lowest energy structures in the Prolactin and Rhodopsin simulations were carried out.

\begin{figure}%
    \centering
    \subfloat[Energy-scoring during folding stage.]{
        {\includegraphics[width=0.4\textwidth]{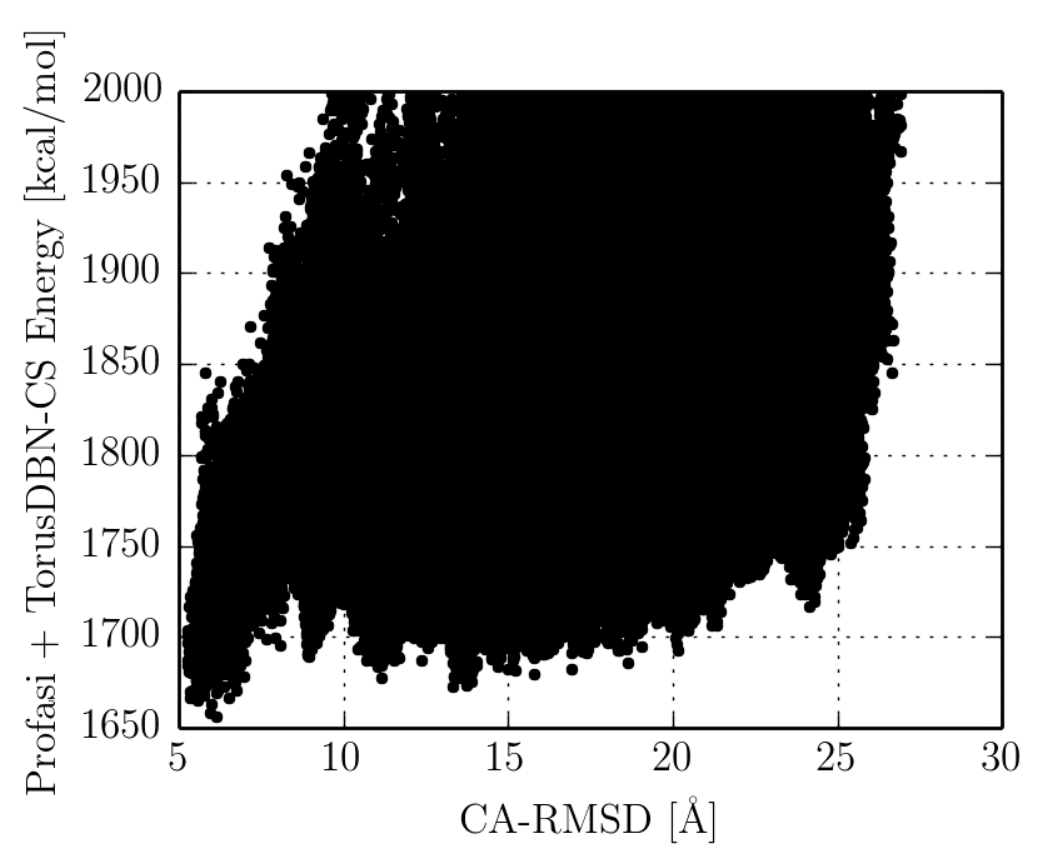}}
    }\quad
    \subfloat[Energy-scoring during refinement stage.]{
        {\includegraphics[width=0.4\textwidth]{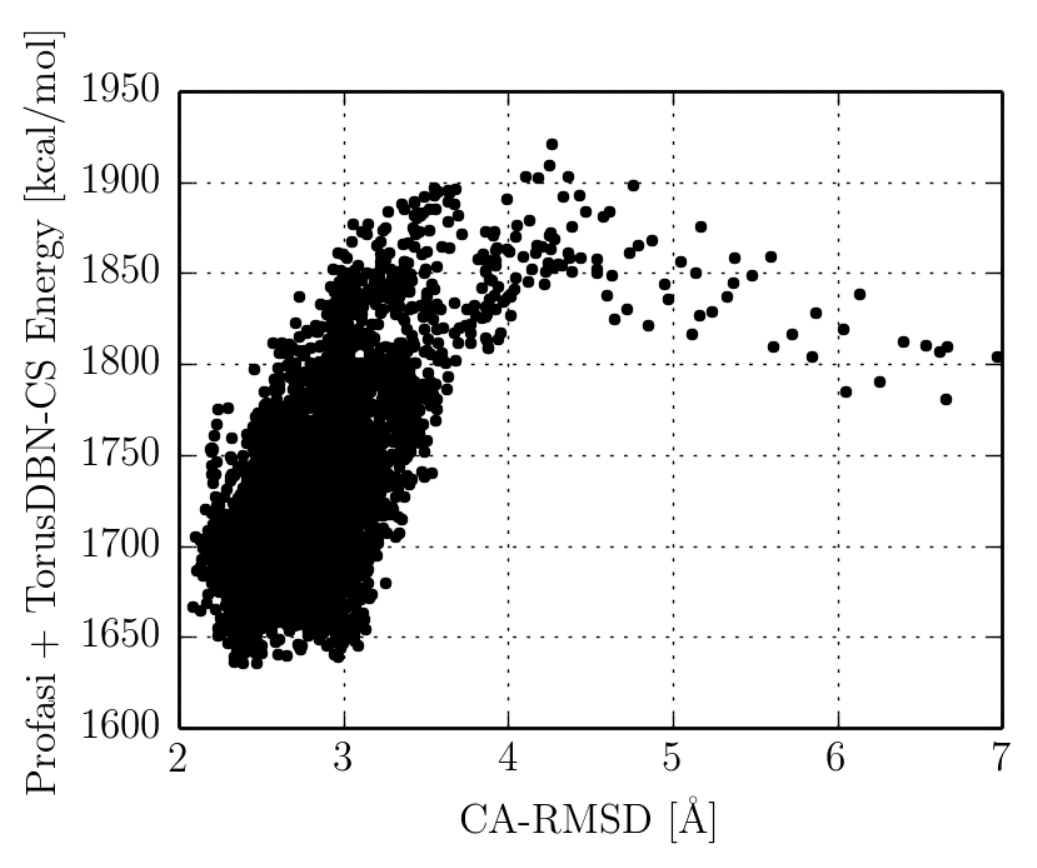}}
    }\\
    \subfloat[Folding stage lowest energy sample (blue). 7.8 \AA~CA-RMSD.]{
        {\includegraphics[width=0.4\textwidth]{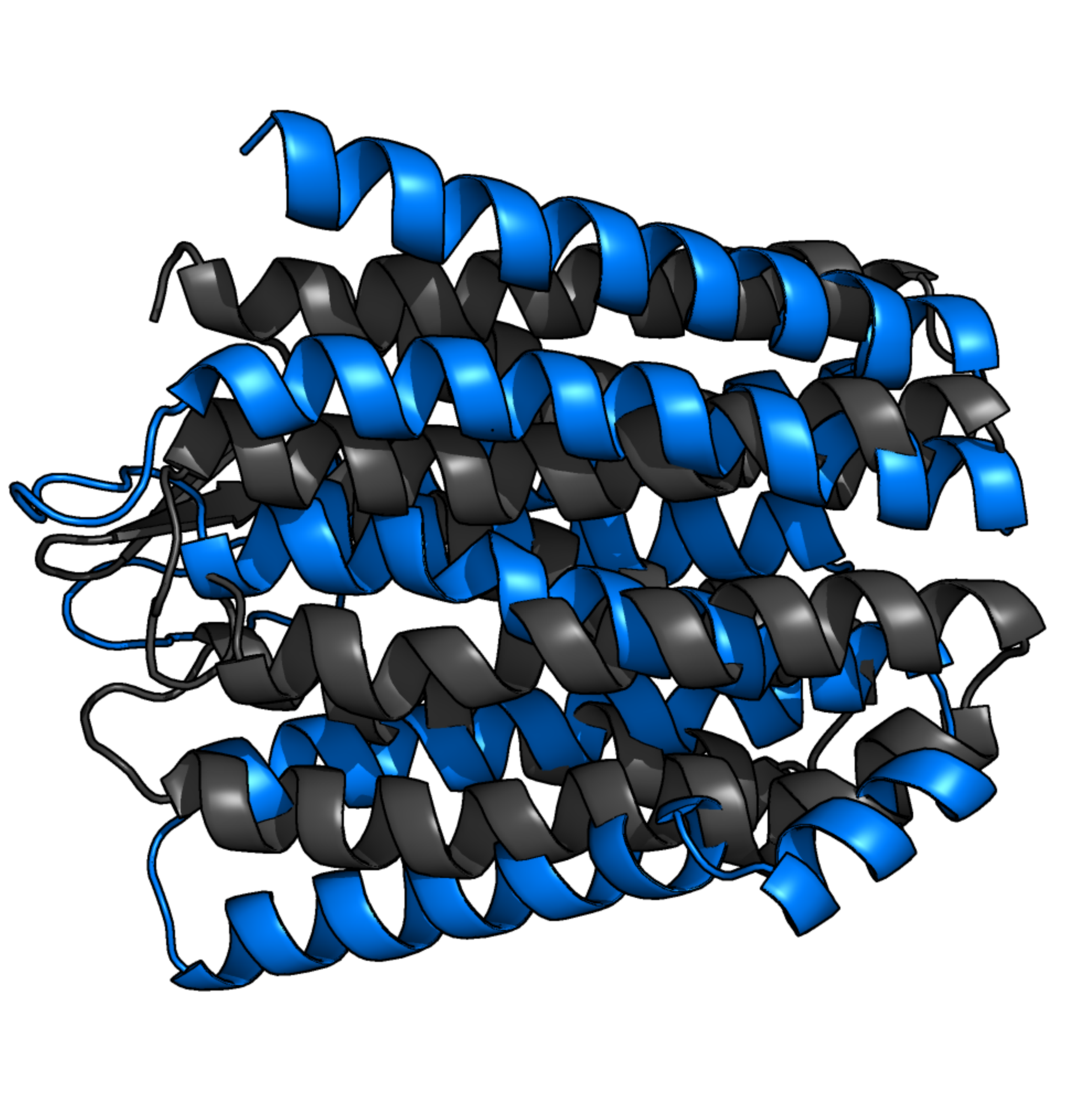}}
    }\quad
    \subfloat[Refinement stage lowest energy sample (blue). 2.5 \AA~CA-RMSD. ]{
        {\includegraphics[width=0.4\textwidth]{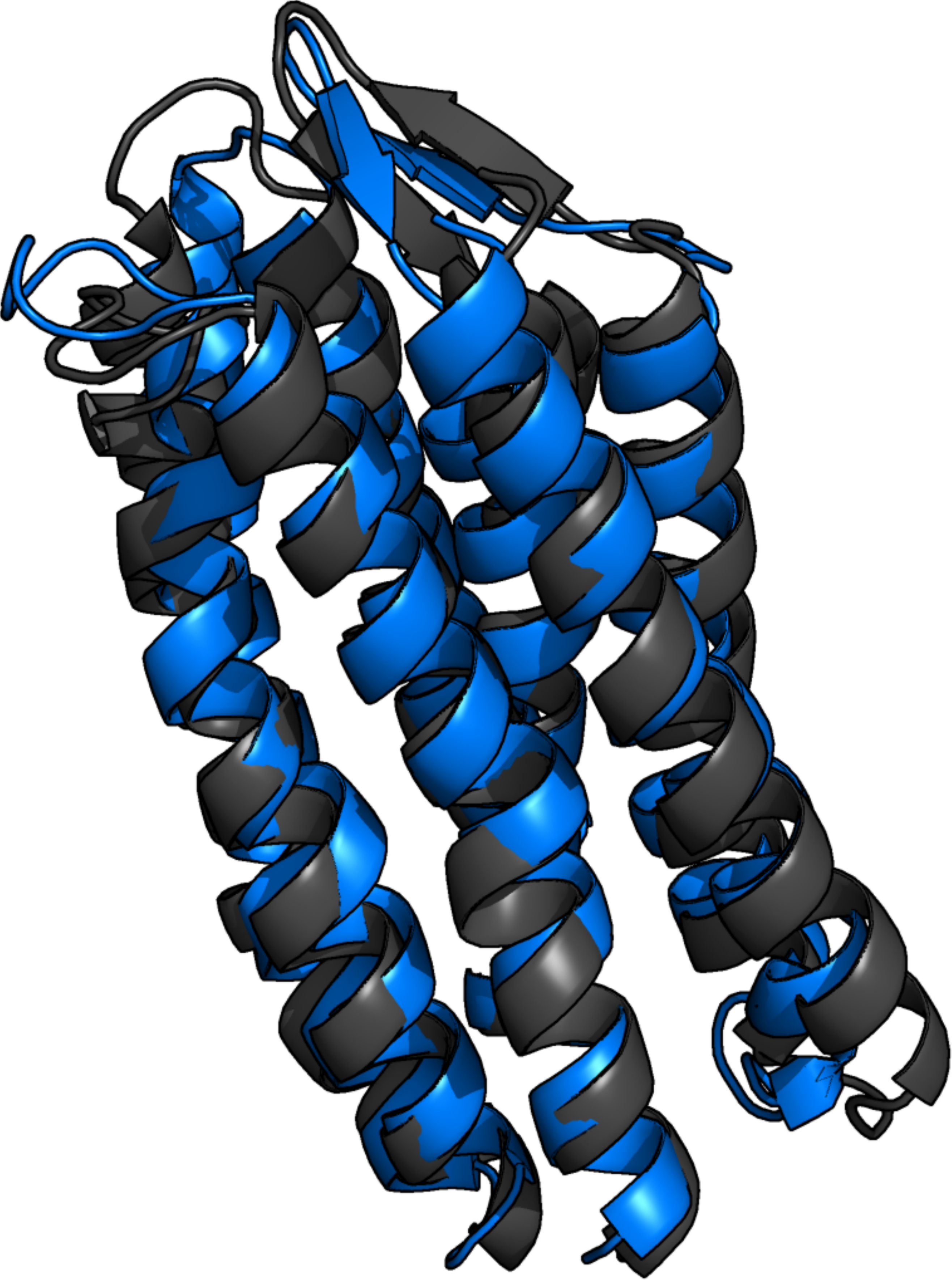}}
    }\\
    \caption{(a) displays the energy scoring during the folding stage of Rhodopsin, and (b) the same statistics during the refinement stage. (c) displays the lowest energy structure after the folding stage, and (d) the lowest energy structure after the refinement stage.}
    \label{fig:rhodopsin}%
\end{figure}

The new module proved very efficient in further minimizing the energy.
Fig.~\ref{fig:rhodopsin} shows the resulting structures and energy/RMSD landscapes from the refinements and folding simulations on Rhodopsin.
The final RMSD after refinement was 2.5 \AA~for Rhodopsin, compared to 7.8 \AA~before refinement.
For Prolactin, the same numbers were 3.5 \AA~and 8.5 \AA, respectively.
The reason for the higher RMSD for Prolactin, compared to Rhodopsin is a flexible handle with no NOE restraints.
The structure of this handle is thus determined by the PROFASI force field and TorusDBN-CS, which apparently does not agree well with the experimental structure in this case - this can be seen from Fig.~\ref{fig:prolactin}.
\\\\The command line to run the refinement is given below:
\begin{lstlisting}
./phaistos --pdb-file rhodopsin_lowest_energy1.pdb \
  --init-from-pdb 1 \
  --iterations 5000000 \
  --threads 4 \
  --monte-carlo-muninn 1 \
  --monte-carlo-muninn-min-beta 0.6 \
  --monte-carlo-muninn-max-beta 1.1 \
  --monte-carlo-muninn-independent-threads 1 \
  --monte-carlo-muninn-weight-scheme multicanonical \
  --monte-carlo-muninn-weight-scheme-use-energy2 1 \
  --backbone-dbn-torus-cs 1 \
  --backbone-dbn-torus-cs-initial-nmr-star-filename \
                                      rhodopsin.str \
  --energy-profasi-cached 1 \
  --energy2-noe 1
  --energy2-noe-active-restraints 140
  --energy2-noe-seamless 1
  --energy2-noe-contact-map-filename noe_ilv.txt
  --move-none 1\
  --move-none-weight 0.005 \
  --move-crisp-dbn-eh 1 \
  --move-crisp-dbn-eh-weight 0.5 \
  --move-semilocal-dbn-eh 1 \
  --move-semilocal-dbn-eh-weight 0.25 \
  --move-sidechain-rotamer 1 \
  --move-sidechain-rotamer-weight 0.25
\end{lstlisting}

\begin{figure}%
    \centering
    \includegraphics[width=\textwidth]{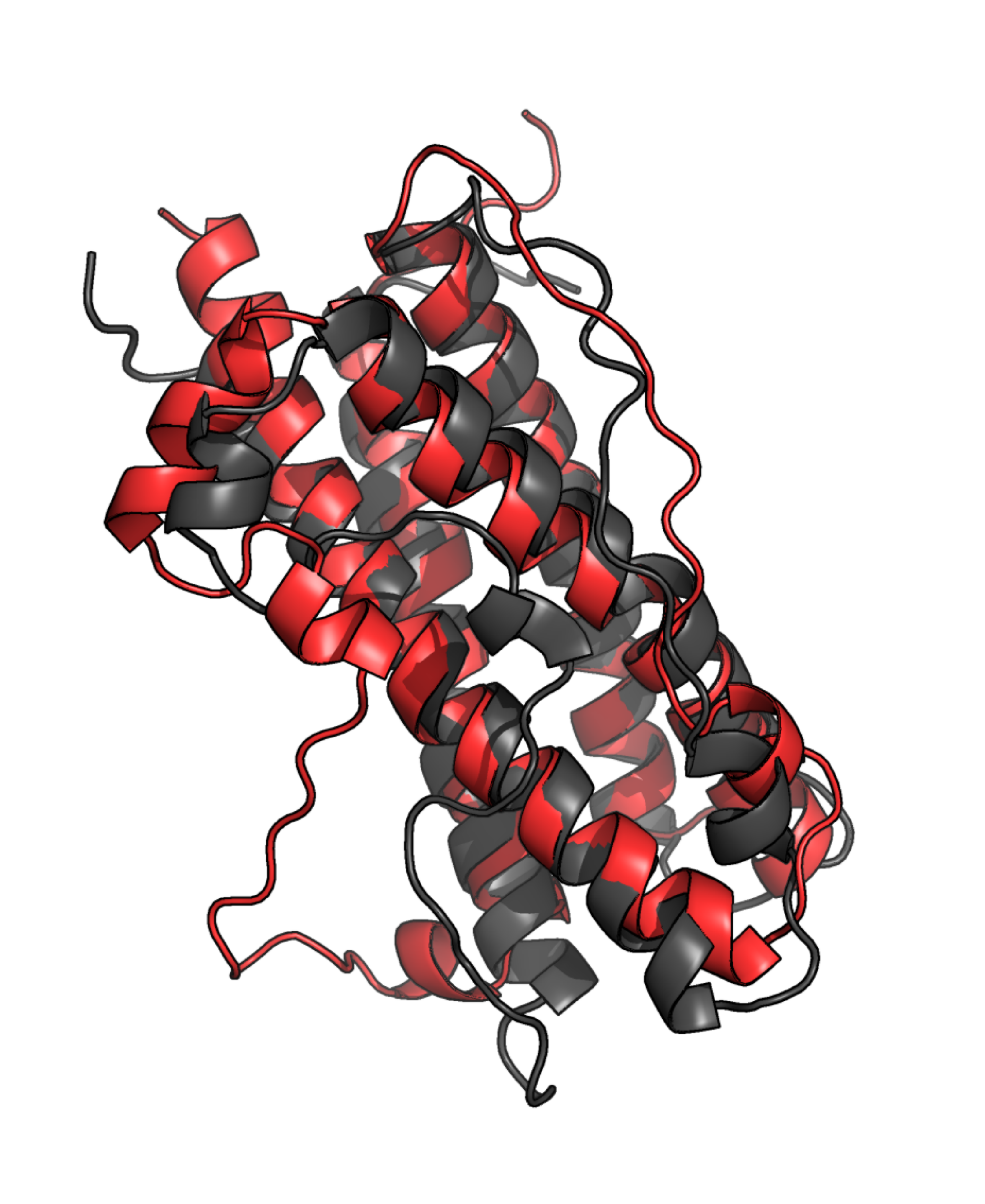}
    \caption{The lowest energy sample (red) for Prolactin after refinement. Note the flexible part which is not in agreement with the experimental X-ray structure (grey).}
    \label{fig:prolactin}%
\end{figure}

\clearpage

\section{Evolutionary distance constraints}

\begin{figure}%
    \centering
    \subfloat[Lowest RMSD sample]{
        \includegraphics[width=0.40\textwidth]{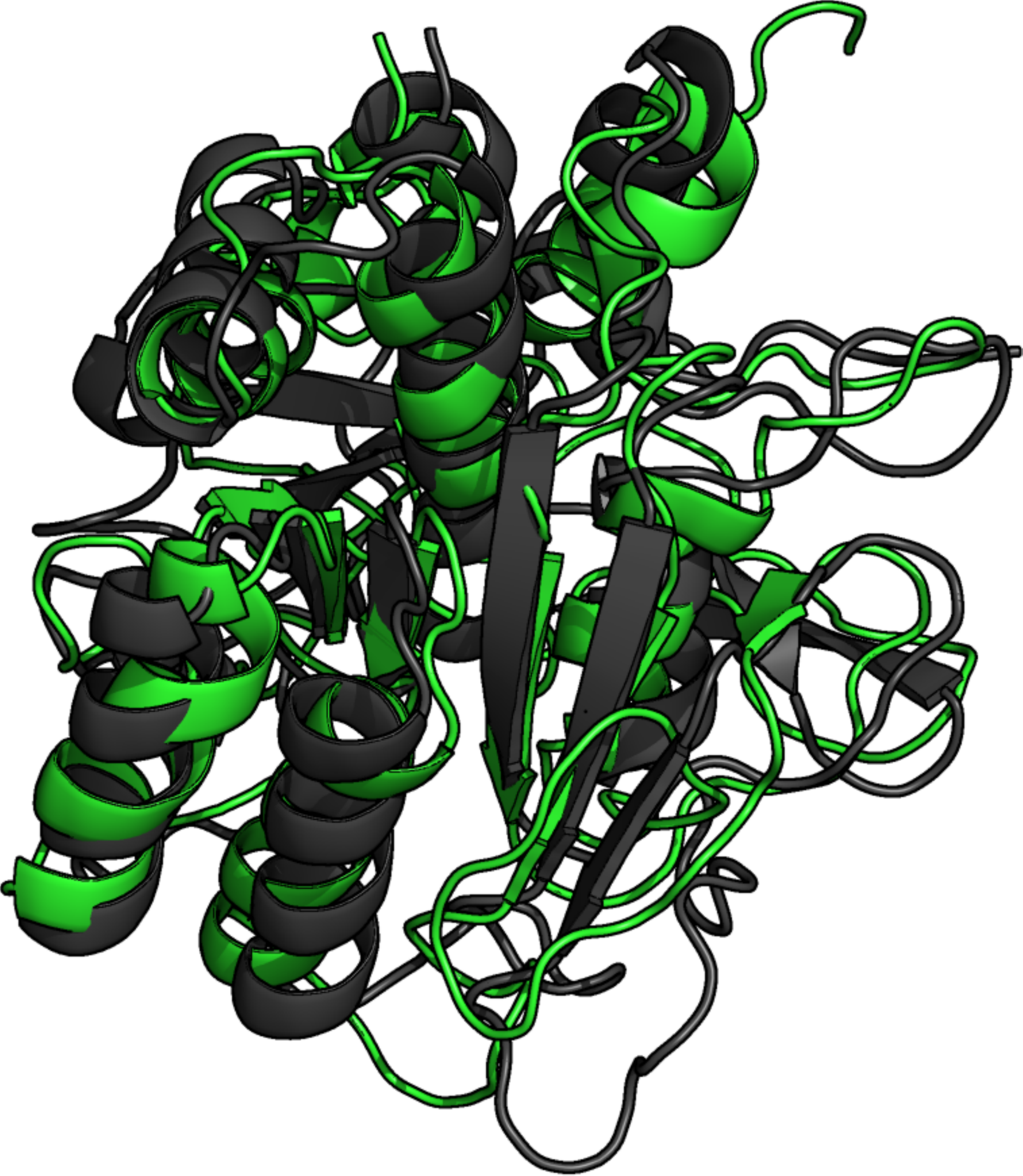}
    }
    \subfloat[Lowest energy sample]{
        \includegraphics[width=0.60\textwidth]{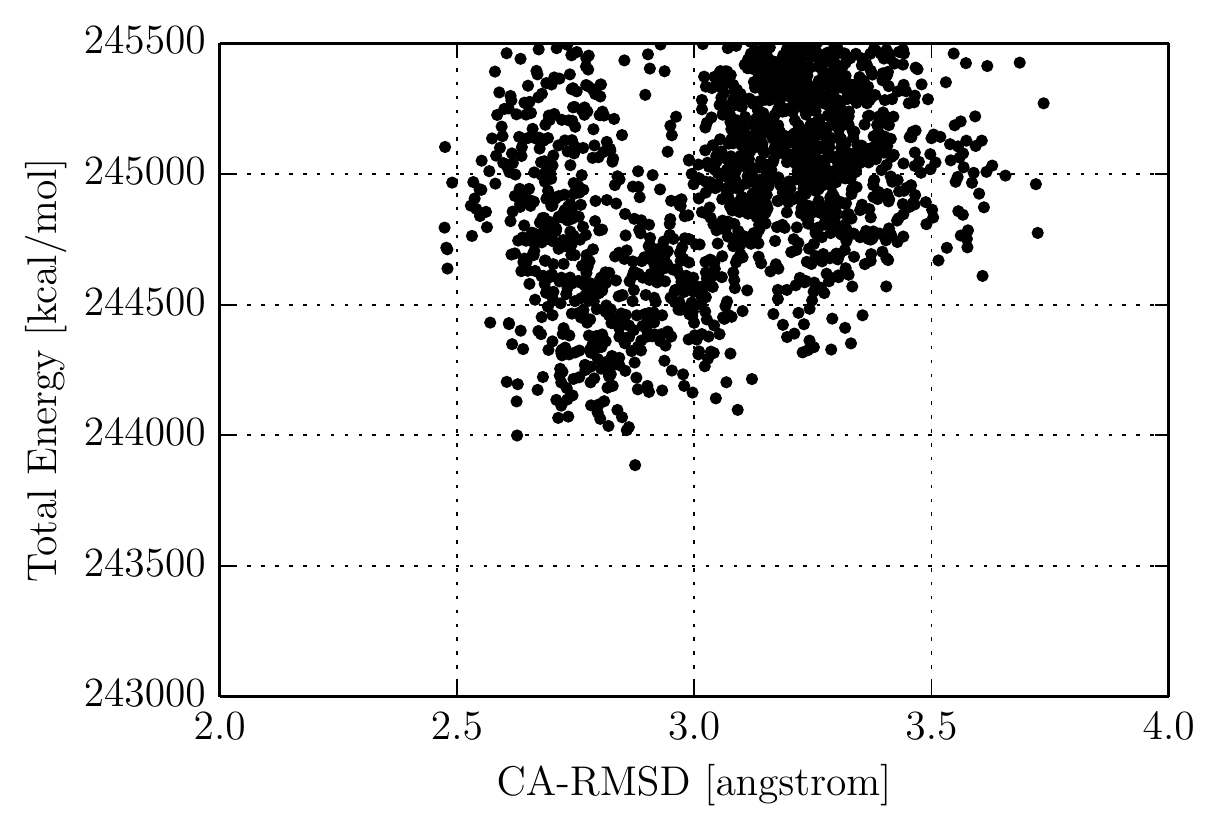}
    }
    \caption{Refinement stage of the savinase simulation. The lowest energy sample has a CA-RMSD of 2.9 \AA. }
    \label{fig:savinase_fold}%
\end{figure}

\begin{figure}
    \includegraphics[width=0.85\textwidth]{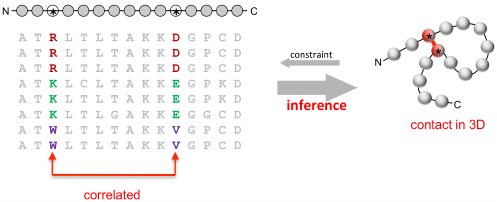}
    \caption{Brief overview of the process from which evolutionary constraints are inferred. Correlated sequence variation that retains favorable interactions is identified and converted to distance constraints. Figure from Marks et al., 2011.}
    \label{fig:evo_constraint}
\end{figure}

As discussed previously, it is increasingly difficult to obtain sufficient distance restraints as the size of the protein increases.
A recently developed methodology uses sequence analysis to infer residue contacts in 3D space \cite{evfold}.
In brief, the method works by identifying sequence co-variation, which retains favorable contacts between residues.
This way, pair of residues which are probable to be close in 3D space can be identified.
The procedure is briefly summarized in Fig.~\ref{fig:evo_constraint}, and is implemented in the EVfold program.

In this proof-of-concept study, 270 contacts were obtained a multiple-sequence alignment using the EVfold program (Wouter Boomsma, personal communications) for the 269 residue protein Savinase.
The restraints were simply treated as NOE restraints using the old NOE code mentioned in the previous section.
A similar simulation to that which folded Rhodopsin was adopted. 
In terms of computational resources, these were increased to 100 threads and $75 \times 10^6$ iterations, compared to only 72 threads and $50 \times 10^6$ iterations for the Rhodopsin simulation.
One thread identified a native-like structure.

The folding simulation yielded a lowest energy structure around 7.5 \AA~CA-RMSD from native. A further refinement with the new NOE code from this structure, yielded a lowest RMSD structure at 2.9 \AA~CA-RMSD from the X-ray structure.
The structure and an energy/RMSD plot for the refinement is shown in Fig.~\ref{fig:savinase_fold}.

%
%

\chapter{Conclusion and Outlook}

During the project described in this thesis and the attached papers, I have implemented a method to determine the structure of several small proteins using their experimental chemical shifts.
The structure of the CI-2 protein was solved rapidly, using only computer resources that are available in any lab, with only chemical shift data that was automatically recorded and assigned.

Lastly, I have attempted to fold several protein structures around 200 amino acids.
Out of 8 proteins greater than 100 residues, a good structure was located in four cases, out of which two were larger than 200 residues.
The last four likely failed due to inefficient use of the NOE restraints.
Since the existing code to handle NOE restraints in PHAISTOS did not perform well on large structures, I implemented a new NOE energy term, and this was used to fold the Rhodopsin structure (225 amino acids) to a CA-RMSD of 2.5 \AA~from the experimental X-ray structure using a set only 195 NOE restraints and assigned backbone chemical shifts, NMR data which had been assigned through automated processes.
The same code was able to fold the Savinase structure (269 amino acids) to a CA-RMSD of 2.9 \AA~from the experimental X-ray structure using only distance restraints derived from evolutionary data and assigned chemical shifts.

This required implementing a version of CamShift, from scratch, in PHAISTOS, and implemented useful energy function rigorously founded in Bayesian statistics.
To aid the setup of calculations, a graphical user interface for PHAISTOS was created.
I have parametrized and implemented a version of ProCS to calculate amide proton chemical shifts, and shown that this parametrization yields structure that are in better agreement with experimental data than simulations using a chemical shift predictor parametrized from experimental data.
Furthermore, I have parametrized parts of the backbone atom ProCS chemical shift predictor and implemented this in a PHAISTOS module.
This required the implementation of FragBuilder Python API which was used to automatically setup, run, and collect data from more than 2,000,000 QM calculations.

The speed of the cached version of the backbone atom ProCS chemical shift predictor will allow an energy function based on chemical shifts to be included in simulations on proteins $>200$ residues. Based on results obtained on the ENHD, Protein G and CI-2 proteins, this will dramatically increase the accuracy of the energy functions that can be used to determine protein structures.

A newly developed NOE energy function shows encouraging results on folding of large structures, and further development of this module is promising.

In conclusion, I have demonstrated, that chemical shifts and sparse NOE data can, in some cases, be used with higher computational efficiency in PHAISTOS, than any other competing method. I have determined a protein structure in less than two days using automatically collected chemical shift data with computational resource available to any lab.
Lastly, I have folded some of the larges protein structures ever folded using similar approaches, while using only modest amount computational resources, compared to current \textit{state-of-the-art} methods.

\bibliography{references}

\end{document}